\documentclass[reprint,onecolumn,english,nofootinbib,letterpaper,nobalancelastpage,superscriptaddress]{revtex4-1}

\usepackage[utf8x]{inputenc}
\usepackage{amsmath}
\usepackage{amssymb}
\usepackage{bbm}
\usepackage{babel}
\usepackage{graphicx}
\usepackage{verbatim}
\usepackage{latexsym}
\usepackage{mathptm}
\usepackage{color}
\usepackage{slashed}

\newcommand{\der}[2]{\frac{\partial #1}{\partial #2}}

\begin{document}

\title{Causality and momentum conservation from relative locality}

\author{{\bf Giovanni AMELINO-CAMELIA}}
\affiliation{Dipartimento di Fisica, Università “La Sapienza” P.le A. Moro 2, 00185 Roma, Italy}
\affiliation{Sez.~Roma1 INFN, P.le A. Moro 2, 00185 Roma, Italy}

\author{{\bf Stefano BIANCO}}
\affiliation{Dipartimento di Fisica, Università “La Sapienza” P.le A. Moro 2, 00185 Roma, Italy}
\affiliation{Sez.~Roma1 INFN, P.le A. Moro 2, 00185 Roma, Italy}

\author{{\bf Francesco BRIGHENTI}}
\affiliation{Dipartimento di Fisica e Astronomia dell'Università di Bologna, Via Irnerio 46, 40126 Bologna, Italy}
\affiliation{Sez.~Bologna INFN,  Via Irnerio 46, 40126 Bologna, Italy}

\author{{\bf Riccardo Junior BUONOCORE}}
\affiliation{Dipartimento di Fisica, Università “La Sapienza” P.le A. Moro 2, 00185 Roma, Italy}
\affiliation{Department of Mathematics, King's College London, The Strand, London, WC2R 2LS, UK}

\begin{abstract}
Theories with a curved momentum space, which became recently of interest in the quantum-gravity literature, can in general violate many apparently robust aspects of our current description of the laws of physics, including relativistic invariance, locality, causality and global momentum conservation. We here explore some aspects of the particularly severe pathologies arising in generic theories with curved momentum space for what concerns causality and momentum conservation. However, we also report results suggesting that when momentum space is
maximally symmetric, and the theory is formulated (DSR-)relativistically, with the associated relativity of spacetime locality,
momentum is globally conserved and there is no violation of causality.
\end{abstract}

\maketitle

\newpage

\section{Introduction}
Over the last decade several independent arguments suggested that
the Planck scale might characterize a non-trivial geometry of momentum space (see, {\it e.g.}, Refs.~\cite{majidCURVATURE,dsr1,dsr2,jurekDSMOMENTUM,girelliCURVATURE,schullerCURVATURE,changMINIC,principle}).
Among the reasons of interest in this possibility we should mention
 approaches to the study of the quantum-gravity problem based on
 spacetime noncommutativity, particularly when considering models with ``Lie-algebra spacetime noncommutativity", $[x_\mu , x_\nu]= i \zeta^\sigma_{\mu \nu} x_\sigma$, where
 the momentum space on which spacetime coordinates generate translations is evidently curved (see, {\it e.g.}, Ref.~\cite{gacmaj}).
 Also in the Loop Quantum Gravity approach~\cite{rovelliLRR} one can adopt a perspective suggesting momentum-space curvature (see, {\it e.g.}, Ref.~\cite{leeCURVEDMOMENTUM}).
 And one should take notice of the fact that the only quantum gravity we actually know
 how to solve,
quantum gravity in the 2+1-dimensional case, definitely does predict a curved momentum space (see, {\it e.g.},
Refs.~\cite{matschull,BaisMullerSchroers,dsr3FREIDLIVINE,bernschr2012,stefanoCQG2013}).

In light of these findings it is then important to understand what are the implications
of curvature of momentum space. Of course the most promising avenue is the one of accommodating this new
structure while preserving to the largest extent possible the structure of our current theories. And some progress
along this direction has already been made in works adopting the  ``relative-locality curved-momentum-space
framework", which was recently proposed in Ref.\cite{principle}.
Working within this framework it was in particular shown~\cite{GiuliaFlavio,GACarXiv11105081,cortes}
that some theories on curved momentum spaces can be formulated
as relativistic theories. These are not
special-relativistic theories, but they are  relativistic
 within the scopes of the proposal
of {}``DSR-relativistic theories"~\cite{dsr1,dsr2}
(also see Refs.~\cite{jurekDSRfirst,leejoaoPRDdsr,leejoaoCQGrainbow,jurekDSRreview,gacSYMMETRYreview}),
theories with two relativistic invariants, the speed-of-light scale
$c$ and a length/inverse-momentum scale $\ell$: the scale that characterizes
the geometry of momentum space must in fact be an invariant if the
theories on such momentum spaces are to be relativistic.

For what concerns locality some works based on Ref.\cite{principle}
have established that, while for generic theories on curved momentum spaces locality is simply lost,
in some appropriate cases the curvature of momentum space
is compatible with only a relatively mild weakening of locality. This is the notion
of relative spacetime locality, such that \cite{whataboutbob} events observed as coincident by nearby observers
may be described as non-coincident by some distant observers. In presence of relative spacetime locality one can
still enforce as a postulate that physical processes are local, but needing the additional specification
that they be local for nearby observers.

The emerging assumption is that research in this area should give priority to theories on curved momentum space
which are (DSR-)relativistic and have relative locality. Of course, it is important to establish whether
these two specifications are sufficient for obtaining acceptable theories. Here acceptable evidently means
theories whose departures from current laws are either absent or small enough to be compatible with
the experimental accuracy with which such laws have been so far confirmed experimentally.
In this respect some noteworthy potential challenges have been exposed in the recent studies in Ref.\cite{linq}
and in Ref.\cite{andrb}. Ref.\cite{linq} observed that, in general, theories on curved momentum space do not
preserve causality, whereas Ref.\cite{andrb} observed that, in general, theories on curved momentum space, even when one enforces
momentum conservation at interactions, may end up loosing global momentum conservation.

The study we here report intends to contribute to the understanding
of theories formulated in the relative-locality curved-momentum-space
framework proposed in Ref.\cite{principle}.
Like Refs.\cite{linq,andrb} we keep our analysis explicit by focusing on the case of the so-called $\kappa$-momentum
space, which is known
to be compatible with a (DSR-)relativistic formulation of theories.
Our main focus then is on establishing whether enforcing relative locality is sufficient for addressing
the concerns for causality
reported in Ref.\cite{linq} and the concerns for momentum conservation reported in
Ref.\cite{andrb}. This is indeed what we find: enforcing relative locality for theories on the $\kappa$-momentum
space is sufficient for excluding the causality-violating processes of Ref.\cite{linq} and the
processes violating global momentum conservation of
Ref.\cite{andrb}.
 And we find further motivation for adopting a DSR-relativistic setup,
 with relative locality, by showing that instead for a generic curved momentum space (non-relativistic, without relative locality) the violations of causality are even more severe than previously established.

A key role in our analysis is played by translation transformations in relativistic theories with a curved momentum space.
As established in previous works~\cite{anatomy,cortes} the relevant laws of translation transformations are in some sense
less rigid than in the standard flat-momentum-space case, but still
must ensure that all interactions are local as described
by nearby observers. It is of course only through such translation transformations that one can enforce relative spacetime locality
for chains of events such as those considered in Refs.\cite{linq,andrb}. In presence of a chain of events any given observer is at most ``near"
one of the events (meaning that the event occurs in the origin of the observer's reference frame) and, because of relative locality,
that observer is then not in position to establish whether or not other events in the chain are local. Enforcing the principle of relative
locality~\cite{principle}
then requires the use of translation transformations connecting at least as many observers as there are distant events in the chain:
this is the only way for enforcing
the spacetime locality of each event in the chain, in the sense of the principle of relative locality.

The main issues and structures we are here concerned with are already fully active and relevant in
the case of $1+1$ spacetime dimensions and at {leading order} in the scale $\ell$ of curvature of momentum space.
We shall therefore mainly focus on the 1+1-dimensional case and on {leading-in-$\ell$-order} results, so that our derivations
can be streamlined a bit and the conceptual aspects are more easily discussed.

\section{Preliminaries on classical particle theories on the $\kappa$-momentum space}\label{digitalsec}
As announced our analysis adopts the
relative-locality curved-momentum-space
framework proposed in Ref.\cite{principle}, and for definiteness focuses on the $\kappa$-momentum space.
This $\kappa$-momentum space is based on a form of on-shellness and a form of the law
of composition of momenta inspired by the $k$-Poincar\'e Hopf algebra \cite{majrue,lukieANNALS}, which had already been of interest from
the quantum-gravity perspective for independent reasons \cite{gacmaj,dsr3FREIDLIVINE,leeCURVEDMOMENTUM}.
The main characteristics of this momentum space are that, at leading order in the deformation scale $\ell$,
 the on-shellness of a particle of momentum $p$ and mass $m_p$ is
\begin{equation} \label{geomassJ}
\mathcal C_p \equiv p_0^2-p_1^2 -\ell p_0 p_1^2 = m^2_p \ ,
\end{equation}
while the composition of two momenta $p$, $q$ is
\begin{equation}
\begin{split}
(p \oplus q)_0 &= p_0 + q_0 \ ,\\
(p \oplus q)_1 &= p_1 + q_1 -\ell p_0 q_1 \ .
\end{split}
\label{jjj}
\end{equation}
Useful for several steps of the sort of analyses we are here interested in is the introduction
of the ``antipode" of the composition law, denoted by $\ominus$, such that $(q \oplus (\ominus q))_\mu =0 = ((\ominus q) \oplus q)_\mu$.
For the $\kappa$-momentum case one has that
$$(\ominus q)_0 = - q_0 ~,~~~(\ominus q)_1 = - q_1 - \ell q_0 q_1$$

We shall not review here the line of analysis which describes these rules of kinematics
as the result of adopting on momentum space the de Sitter metric and a specific torsionful affine connection. These
points are discussed in detail in Refs.~\cite{GiuliaFlavio,anatomy}.

In light of our objectives it is useful for us to briefly summarize here the description of
events within the relative-locality curved-momentum-space
framework. More detailed and general discussions of this aspect can be found
in Refs.~\cite{principle,anatomy}. Here we shall be satisfied  with briefly describing the
illustrative case of the event
in Fig.1, for which we might  think for example of the event of
absorption of a photon by an atom.
The case of interest in the recent literature on the relative-locality framework is the one of events of this sort analyzed within classical mechanics (so, in particular, the diagram shown here in Fig.1 should not be interpreted in the sense of quantum theory's Feynman diagrams, bur rather merely as a schematic description of a classical-physics event).
\begin{figure}[h!]
\begin{center}
\includegraphics[scale=0.5]{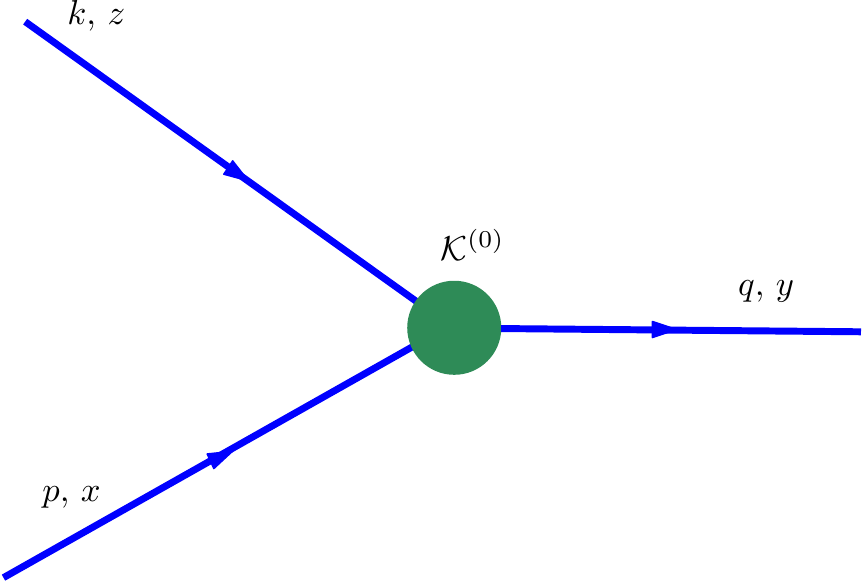}
\end{center}
\caption{{\small We here show schematically a 3-valent event marked by a $\mathcal{K}^{(0)}$ that symbolizes a boundary term conventionally located at value  $s_0$ of the affine parameter  $s$. The boundary term enforces (deformed)\ momentum conservation at the event.}}
\label{vertex}
\end{figure}

The formalism introduced in Ref.~\cite{principle} allows the description of such an event in terms
of the law of on-shellness, which for the $\kappa$-momentum space is (\ref{geomassJ}),
and the law of composition of momenta, which for the $\kappa$-momentum space is (\ref{jjj}).
This is done by introducing the action~\cite{principle}

\begin{equation}\label{actionj}
\begin{array}{lll}
\mathcal{S}
&=&
\int_{-\infty}^{s_{0}}ds\left(z^{\mu}\dot{k}_{\mu}+\mathcal{N}_k [\mathcal{C}_{k} - m_k^2]\right)+
\int_{-\infty}^{s_{0}}ds\left({x}^{\mu}\dot{p}_{\mu}+\mathcal{N}_p[\mathcal{C}_{p} - m_p^2]\right)
+\int_{s_{0}}^{+\infty}ds\left(y^\mu\dot q_{\mu}+\mathcal{N}_{q}[\mathcal{C}_{q} - m_q^2]\right)
-\xi_{(0)}^{\mu}\mathcal{K}_{\ \mu}^{(0)}
\, .
\end{array}
\end{equation}
Here the Lagrange multipliers $\mathcal{N}_k$,$\mathcal{N}_p$,$\mathcal{N}_q$
enforce in standard way the on-shellness of particles.
The most innovative part of the formalization introduced
in Ref.~\cite{principle}
is the presence of boundary terms at endpoints of
worldlines, enforcing momentum conservation. In the case of (\ref{actionj}), describing
the single interaction in Fig.1, there is only one such boundary term, and the momentum-conservation-enforcing $\mathcal{K}_{\ \mu}^{(0)}$
takes the form\footnote{Note that for associative composition laws, as is the case of the $\kappa$-momentum-space composition law (\ref{jjj}),
on can rewrite $(k\oplus p)_\mu-q_\mu =0$ equivalently as $((k\oplus p) \oplus (\ominus q))_\mu =0$. This is due to the logical chain
$((k\oplus p) \oplus (\ominus q))_\mu =0 ~ \Rightarrow ((k\oplus p) \oplus (\ominus q) \oplus q)_\mu=q_\mu ~ \Rightarrow
(k\oplus p)_\mu =q_\mu $.}
\begin{equation}
\begin{split}
{\cal K}^{(0)}_{\ \mu}&= (k\oplus p)_\mu-q_\mu \ .
\label{constraints}
\end{split}
\end{equation}

Relative spacetime locality is an inevitable feature of descriptions of events governed by curvature of momentum space
of the type illustrated by our example (\ref{actionj}). To see this we vary the action (\ref{actionj})
keeping the momenta fixed at $s = \pm \infty$, as prescribed in Ref.~\cite{principle},
and we find the equations of motion
\begin{gather}
\dot k_\mu =0~,~~\dot p_\mu =0~,~~\dot q_\mu =0~,~~\\
\mathcal{C}_k=m_k^2~,~~\mathcal{C}_p=m_p^2~,~~\mathcal{C}_q=m_q^2~,~~\\
\mathcal{K}_\mu^{(0)}=0,\\
\dot z^\mu = \mathcal{N}_k \frac{\partial \mathcal{C}_k}{\partial k_\mu}~,~~~
\dot x^\mu = \mathcal{N}_p \frac{\partial \mathcal{C}_p}{\partial p_\mu}~,~~~
\dot y^\mu = \mathcal{N}_q \frac{\partial \mathcal{C}_q}{\partial q_\mu} \, ,
\label{jjjbbb}
\end{gather}
and the boundary conditions at the endpoints of the 3 semi-infinite worldlines
\begin{equation}
z^\mu(s_{0}) = \xi^\nu_{(0)} \frac{\partial \mathcal{K}_\nu^{(0)}}{\partial k_\mu}~,~~~
x^\mu(s_{0}) = \xi^\nu_{(0)} \frac{\partial \mathcal{K}_\nu^{(0)}}{\partial p_\mu}~,~~~
y^\mu(s_{0}) =- \xi^\nu_{(0)} \frac{\partial \mathcal{K}_\nu^{(0)}}{\partial q_\mu} \ .
\label{jjjbbbRRR}
\end{equation}

The relative locality is codified in the fact that
for configurations
 such that $\xi_{(0)}^{\mu} \neq 0$ the boundary conditions (\ref{jjjbbbRRR})
impose that the endpoints of the worldlines do not coincide, since in general
\begin{equation}
\frac{\partial \mathcal{K}_\nu^{(0)}}{\partial k_\mu} \neq \frac{\partial \mathcal{K}_\nu^{(0)}}{\partial p_\mu} \neq -\frac{\partial \mathcal{K}_\nu^{(0)}}{\partial q_\mu} \ ,
\end{equation}
so that in the coordinatization of the (in that case, distant)
observer the interaction appears to be non-local.
However, as shown in Fig.2, for observers such that the same configuration is described with $\xi_{(0)}^{\mu} =0$ the endpoints of the worldlines
must coincide and be located in the origin of the observer
($x^\mu(s_{0}) = y^\mu(s_{0}) = z^\mu(s_{0})=0$).
\begin{figure}[h!]
\includegraphics[width= 0.4\columnwidth]{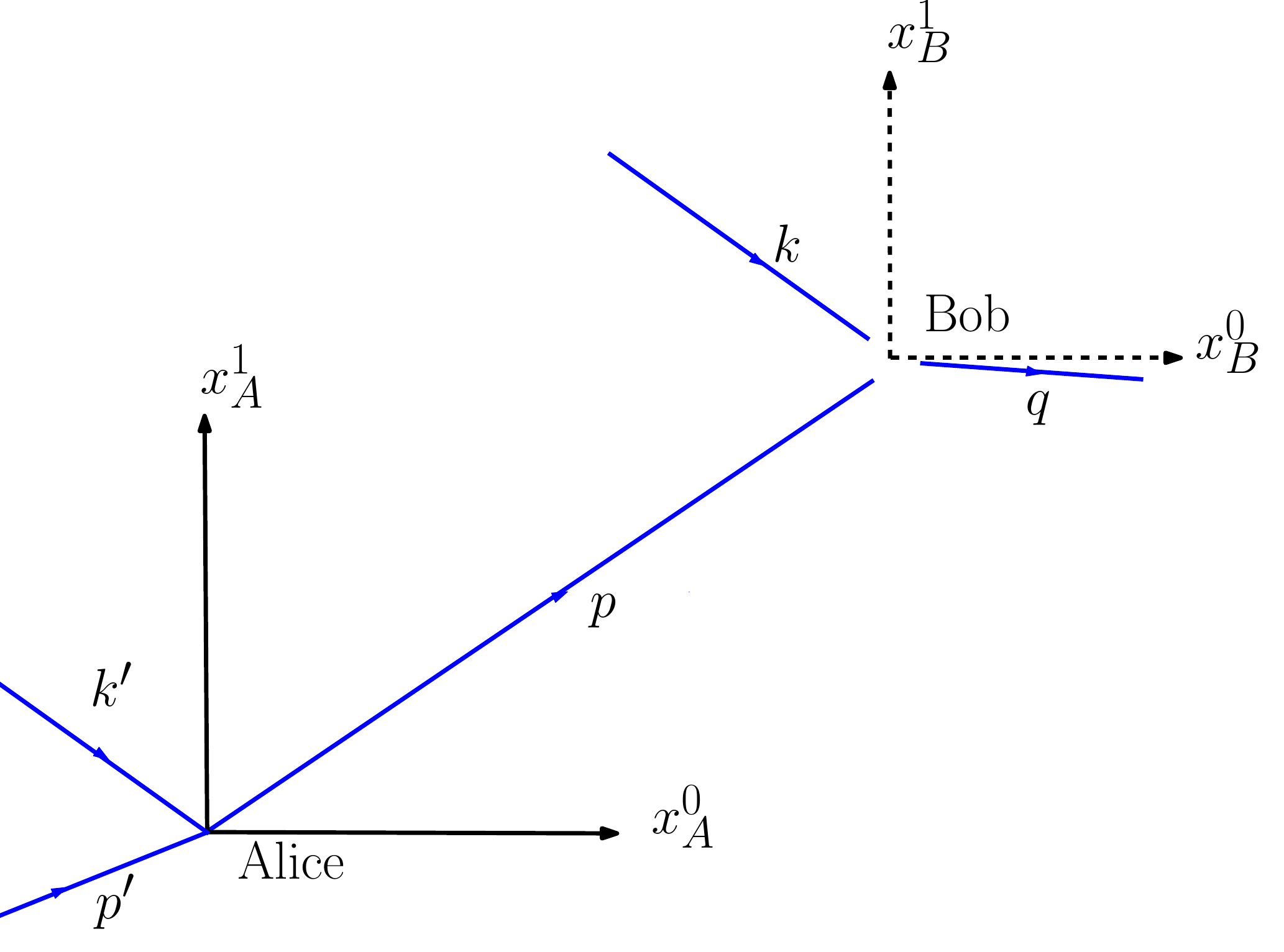}
\hspace{2cm}
\includegraphics[width= 0.4\columnwidth]{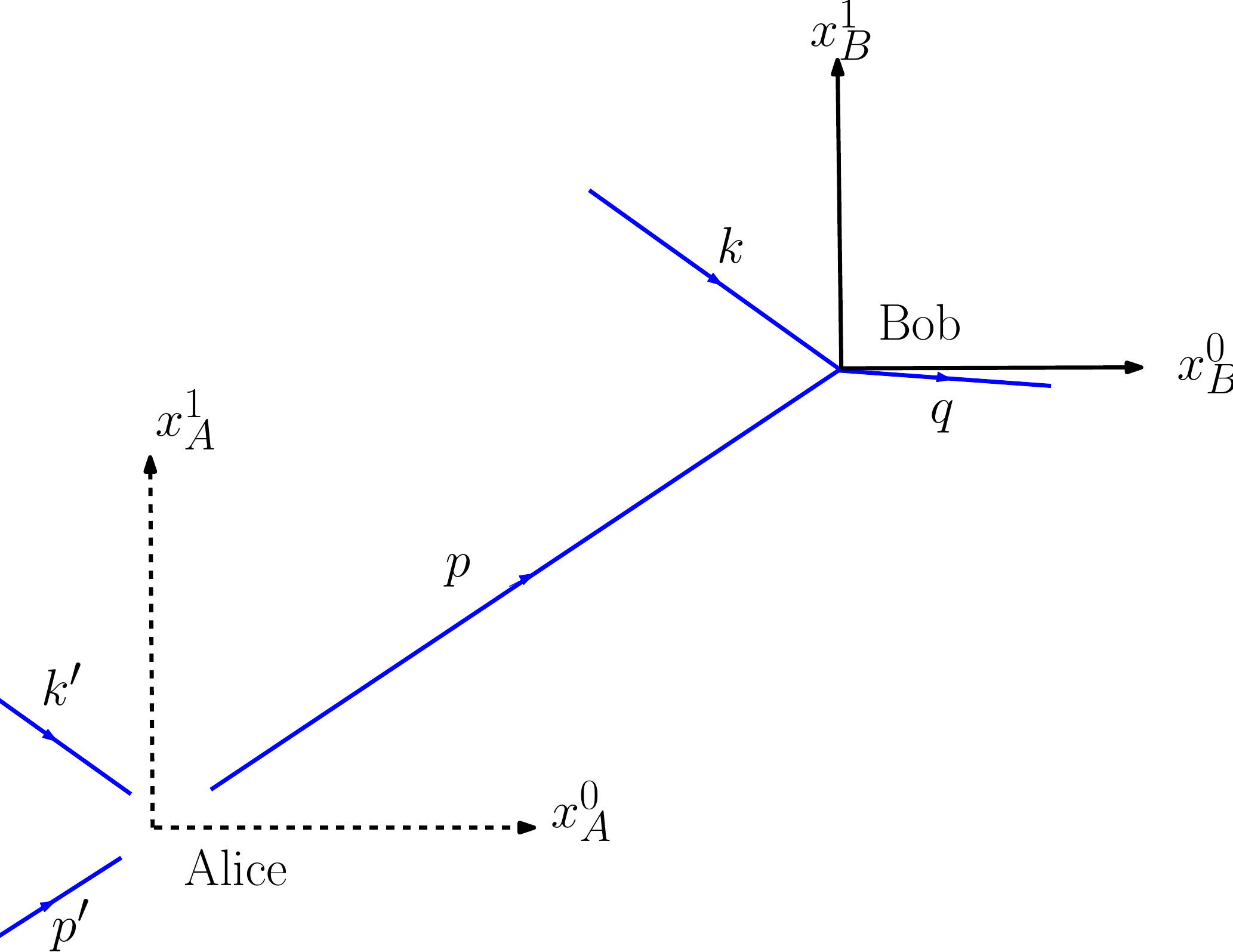}
\caption{ \small{We here give a schematic description of
a process composed of two causally-connected events. The event at Alice could
be the absorption of a photon by an atom and the event at Bob could be another absorption of a photon by the same atom. The implications of relative locality are visualized
by describing Alice's perspective on the process in the left panel and
the perspective of Bob (distant from Alice and in relative rest with respect to Alice)
in the right panel. According to Alice's description  the first absorption event
(which occurs in Alice's origin of the reference frame) is local, but Alice's
inferences about the
second absorption event (which occurs at Bob, far away from Alice) would
characterize it as non-local. Bob has a relativistically specular viewpoint: Bob's description of the second absorption event
(which occurs in Bob's origin of the reference frame) is local but Bob's
inferences about the
first absorption event (which occurs at Alice, far away from Bob) would
characterize it as non-local. This is how a pair of causally-connected distant local events gets described in presence of relative locality.}}
\end{figure}
And it is important to notice that taking as starting point
of the analysis some observer
Alice for whom  $\xi_{(0)[A]}^\mu \neq 0$, {\it i.e.} an observer distant from
the interaction who sees the interaction as non-local,
one can obtain from Alice an observer
Bob for whom $\xi_{(0)[B]}^\mu =0$
if the transformation from Alice to Bob
for endpoints of coordinates
has the form
\begin{equation}
\begin{split}
z^\mu_{B}(s_{0}) &= z^\mu_{A}(s_{0})
- \xi_{A}^\nu \frac{\partial \mathcal{K}_\nu^{(0)}}{\partial k_\mu} \ ,
\\
x^\mu_{B}(s_{0}) &= x^\mu_{A}(s_{0})
- \xi_{A}^\nu \frac{\partial \mathcal{K}_\nu^{(0)}}{\partial p_\mu} \ ,\\
y^\mu_{B}(s_{0}) &= y^\mu_{A} (s_{0})
+ \xi_{A}^\nu \frac{\partial \mathcal{K}_\nu^{(0)}}{\partial q_\mu} \ .
\end{split}
\end{equation}
Such a property for the endpoints is produced of course,
for the choice $b^\nu = \xi_{A}^\nu$,
by  the corresponding prescription for the
 translation transformations:
\begin{equation}
\begin{split}
x^\mu_{B}(s) &= x^\mu_{A}(s) -b^\nu \frac{\partial \mathcal{K}_\nu}{\partial p_\mu}= x^\mu_{A}(s)+b^\nu\{(k\oplus p)_\nu,x^\mu(s)\} \ ,\\
z^\mu_{B}(s) &= z^\mu_{A}(s) -b^\nu \frac{\partial \mathcal{K}_\nu}{\partial k_\mu}=z^\mu_{A}(s)+b^\nu\{(k\oplus p)_\nu,z^\mu(s)\}\ ,\\
y^\mu_{B}(s) &= y^\mu_{A}(s) +b^\nu \frac{\partial \mathcal{K}_\nu}{\partial q_\mu}=y^\mu_{A}(s)+b^\nu\{q_\nu,y^\mu(s)\} \ ,\\
\label{traslprl}
\xi_{B}^\mu &= \xi_{A}^\mu -b^\mu\ .
\end{split}
\end{equation}
where it is understood that $\{x^\mu,p_\nu\}=\delta^\mu_\nu$, $\{z^\mu,k_\nu\}=\delta^\mu_\nu$, $\{y^\mu,q_\nu\}=\delta^\mu_\nu$. This also shows that in this framework one can enforce the ``principle of relative locality" \cite{principle} that all
 interactions
are local according to nearby observers
(observers such that the interaction occurs in the origin of their reference frame).

\section{Cause and effect, with relative locality}\label{digitalsec}
\begin{figure}[h!]
\begin{center}
\includegraphics[scale=0.5]{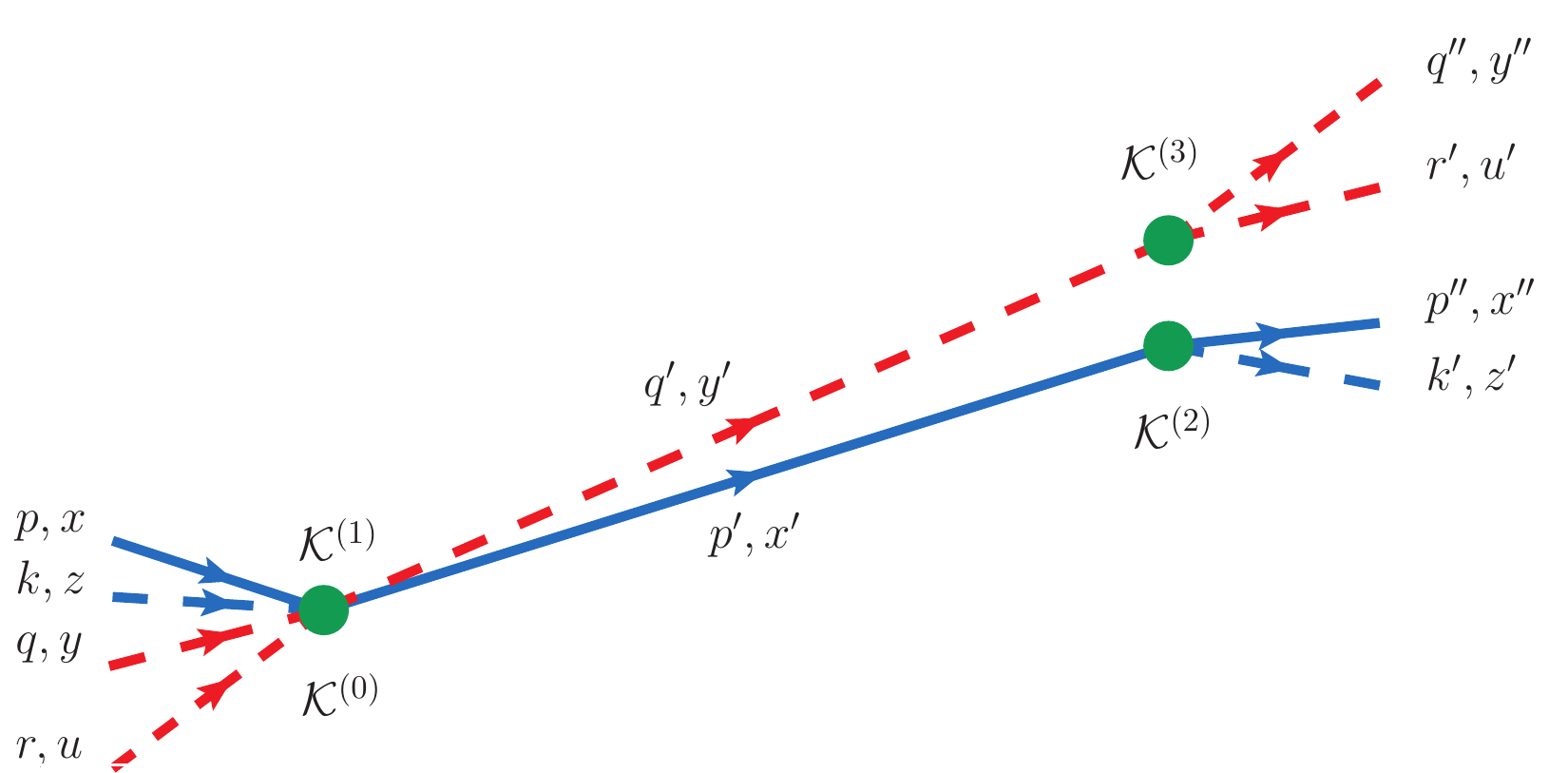}
\end{center}
\caption{{\small We here show schematically two causal links:
a pair of causally-connected events is shown in red
and another pair of causally-connected events is shown in blue,
but there is no causal connection (in spite of the coincidence
of the events ${\cal K}^{(0)}$ and ${\cal K}^{(1)}$) between
events where blue lines cross and events where red lines cross. We analyze
this situation with the simplifying assumption that some of the particles involved (those described by dashed lines) have energies small enough that
the Planck-scale effects here of interest can be safely neglected.}}
\label{process}
\end{figure}

Technically our goal is to work within the framework briefly reviewed in the previous section
(and described in more detail and generality in Refs.\cite{principle,anatomy}),
specifically assuming the laws (\ref{geomassJ}) and (\ref{jjj}) for the $\kappa$-momentum space,
and show that
the concerns for causality
reported in Ref.\cite{linq} and the concerns for momentum conservation reported in
Ref.\cite{andrb} do not apply once the principle of relative locality is enforced.
We start with the causality issue and before considering specifically the concerns
discussed in Ref.\cite{linq} we devote this section to an aside on the relationship between
cause and effect in our framework.
We just intend to show that relative locality, though weaker than ordinary absolute locality,
 is strong enough to ensure
 the objectivity of  the causal link
between a cause and its effect.
An example of situation where this is not {\it a priori} obvious with relative locality is the one in Fig.\ref{process},
where we illustrate schematically
two causal links:
a pair of causally-connected events is shown in red
and another pair of causally-connected events is shown in blue,
but there is no causal connection (in spite of the coincidence
of the events ${\cal K}^{(0)}$ and ${\cal K}^{(1)}$) between
events where blue lines cross and events where red lines cross.
An example of situation of the type shown in Fig.\ref{process}
is the one of two atoms getting both coincidently excited by photon absorption, then both propagating freely and ultimately both getting de-excited by emitting a photon each.

A problem might arise when (as suggested in Fig.3) events on two different causal links happen to be rather
close in spacetime: because of relative locality observers distant from such near-coincident (but uncorrelated) events
might get a sufficiently distorted picture of the events that
the causal links could get confused. We will arrange for
a particularly insightful such situation by the end of this section. And ultimately we shall find that no confusion
about causal links arises if information on the different events is gathered by nearby observers.
Specifically for the situation in Fig.\ref{process} it will be necessary to rely on at least two observers:
an observer Alice near events $\mathcal{K}_{\ \mu}^{(0)}$ and $\mathcal{K}_{\ \mu}^{(1)}$
and an observer Bob near events $\mathcal{K}_{\ \mu}^{(2)}$ and $\mathcal{K}_{\ \mu}^{(3)}$.

We shall do this analysis in detail but making some simplifying assumptions about the energies of the particles
involved. For the particles described by dashed lines in Fig.\ref{process} we assume that they are ``soft"~\cite{anatomy},
{\it i.e.} their energies $E$ are small enough that terms of order $\ell E^2$ are negligible
in comparison to all other energy scales that we shall instead take into account.
The particles described by solid lines in Fig.\ref{process} are instead ``hard", meaning that for them $\ell$ corrections
must be taken into account.
We also adopt the simplification that all particles
are ultrarelativistic, {\it i.e.} for massive particles the mass can
be neglected.

The action describing the situation in Fig.\ref{process} within the relative-locality curved-momentum-space
framework proposed in Ref.~\cite{principle} is
\begin{equation}\label{action}
\begin{array}{lll}
\mathcal{S}
&=&
\displaystyle
\int_{-\infty}^{s_{1}}ds\left(z^{\mu}\dot{k}_{\mu}+\mathcal{N}_k\mathcal{C}_{k}^{(0)}\right)+
\int_{-\infty}^{s_{1}}ds\left({x}^{\mu}\dot{p}_{\mu}+\mathcal{N}_p(\mathcal{C}_{p}-m_p^2)\right)+
\int_{-\infty}^{s_{0}}ds\left(y^{\mu}\dot{q}_{\mu}+\mathcal{N}_q(\mathcal{C}_{q}^{(0)}-m^2_q)\right)+\\
&&
\displaystyle
\int_{-\infty}^{s_{0}}ds\left({u}^{\mu}\dot{r}_{\mu}+\mathcal{N}_r\mathcal{C}_{r}^{(0)}\right)+
\int_{s_{1}}^{s_{2}}ds\left({x'}^{\mu}\dot p'_{\mu}+\mathcal{N}_{p'}(\mathcal{C}_{p'}-m_{p'}^2)\right)
+\int_{s_{0}}^{s_{3}}ds\left(y'^{\mu}\dot q'_{\mu}+\mathcal{N}_{q'}(\mathcal{C}_{q'}^{(0)}-m_{q'}^2)\right)+\\
&&
\displaystyle
\int_{s_{3}}^{+\infty}ds\left(y''^\mu\dot q''_{\mu}+\mathcal{N}_{q''}(\mathcal{C}_{q''}^{(0)}-m^2_{q''})\right)+
\int_{s_{3}}^{+\infty}ds\left({u'}^{\mu}\dot r'_{\mu}+\mathcal{N}_{r'}\mathcal{C}_{r'}^{(0)}\right)+\int_{s_{2}}^{+\infty}ds\left(x''^\mu\dot p''_\mu+\mathcal{N}_{p''}(\mathcal{C}_{p''}-m_{p''}^2)\right)+\\
&&
\displaystyle
\int_{s_{2}}^{+\infty}ds\left(z'^{\mu}\dot{k'}_{\mu}+\mathcal{N}_{k'}\mathcal{C}_{k'}^{(0)}\right)
-\xi_{(0)}^{\mu}\mathcal{K}_{\ \mu}^{(0)}
-\xi_{(1)}^{\mu}\mathcal{K}_{\ \mu}^{(1)}
-\xi_{(2)}^{\mu}\mathcal{K}_{\ \mu}^{(2)}
-\xi_{(3)}^{\mu}\mathcal{K}_{\ \mu}^{(3)}\ ,
\end{array}
\end{equation}
where the ${\cal K}^{(i)}_{\ \mu}$  appearing in the boundary terms enforce the relevant conservation laws
\begin{equation}
\begin{split}
{\cal K}^{(0)}_{\ \mu}&= (r \oplus q)_\mu-q'_\mu \ ,\\
{\cal K}^{(1)}_{\ \mu}&= (k\oplus p)_\mu-p'_\mu \ ,\\
{\cal K}^{(2)}_{\ \mu}&= p'_\mu - (k'\oplus p'')_\mu \ ,\\
{\cal K}^{(3)}_{\ \mu}&= q'_\mu-(r' \oplus q'')_\mu \ .
\label{constraints}
\end{split}
\end{equation}
Several aspects of (\ref{action}) are worth emphasizing. First we notice that the action in (\ref{action}) is just the sum of
two independent pieces, one for each (two-event-)chain of causally-connected events.
For soft particles we codified the on-shellness in terms of $\mathcal{C}^{(0)}_{p}=p_0^2-p_1^2$, while for hard particles
we have $\mathcal C_p \equiv p_0^2-p_1^2 -\ell p_0 p_1^2$, appropriate for the $\kappa$-momentum case. For conceptually clarity massive particles in (\ref{action}) are identifiable indeed because
we write a mass term for them, even though, as announced, we shall assume throughout this section that all particles are ultrarelativistic. Also note that the action (\ref{action}) is not specialized to the case which will be here
of interest from the causality perspective, which is the case of coincidence of the two events ${\cal K}^{(0)}$
and ${\cal K}^{(1)}$: we shall enforce that feature later by essentially focusing on cases such that $\xi^\mu_{(0)}=\xi^\mu_{(1)}$.

By varying the action (\ref{action}), one obtains the following equations of motion
\begin{gather*}
\dot p_\mu =0~,~~\dot q_\mu =0~,~~\dot k_\mu =0~,~~\dot r_\mu =0~,~~\dot p'_\mu =0~,~~\dot q'_\mu =0~,~~\dot p''_\mu =0~,~~\dot q''_\mu =0~,~~\dot k'_\mu =0~,~~\dot r'_\mu =0\ ,\\
\mathcal{C}_p=m_p^2~,~~\mathcal{C}_q^{(0)}=m^2_q~,~~\mathcal{C}_k^{(0)}=0~,~~\mathcal{C}_r^{(0)}=0~,~~\mathcal{C}_{p'}=m_{p'}^2~,~~\mathcal{C}_{q'}^{(0)}=m_{q'}^2~,~~\mathcal{C}_{p''}=m_{p''}^2~,~~\mathcal{C}_{q''}^{(0)}=m^2_{q''}~,~~\mathcal{C}_{k'}^{(0)}=0~,~~\mathcal{C}_{r'}^{(0)}=0\ ,\\
{\cal K}^{(0)}_\mu=0 ~,~~~{\cal K}^{(1)}_\mu=0 ~,~~~{\cal K}^{(2)}_\mu=0 ~,~~~{\cal K}^{(3)}_\mu=0\ ,\\
\dot x^\mu = \mathcal{N}_p \der{\mathcal{C}_p}{p_\mu}~,~~~
\dot y^\mu = \mathcal{N}_q \der{\mathcal{C}_q^{(0)}}{q_\mu}~,~~~
\dot z^\mu = \mathcal{N}_k \der{\mathcal{C}_k^{(0)}}{k_\mu}~,~~~
\dot u^\mu = \mathcal{N}_{r} \der{\mathcal{C}_r^{(0)}}{ r_\mu}~,~~~\\
\dot x'^\mu = \mathcal{N}_{p'} \der{\mathcal{C}_{p'}}{ p'_\mu}~,~~~
{\dot y}'^\mu = \mathcal{N}_{q'} \der{\mathcal{C}_{q'}^{(0)}}{ q_\mu'}~,~~~
\dot x''^\mu = \mathcal{N}_{p''} \der{\mathcal{C}_{p''}}{p''_\mu}~,~~~
\dot y''^\mu = \mathcal{N}_{q''} \der{\mathcal{C}_{q''}^{(0)}}{q''_\mu}~,~~~
\dot z'^\mu = \mathcal{N}_{k'} \der{\mathcal{C}_{k'}^{(0)}}{k'_\mu}~,~~~
\dot u'^\mu = \mathcal{N}_{r'} \der{\mathcal{C}_{r'}^{(0)}}{r'_\mu}~,~~~\\
\end{gather*}
and the boundary conditions for the endpoints of the worldlines
\begin{gather*}
x^\mu(s_{1}) = \xi^\nu_{(1)} \der{\mathcal{K}^{(1)}_\nu}{p_\mu}~,~~~
z^\mu(s_{1}) = \xi^\nu_{(1)} \der{\mathcal{K}^{(1)}_\nu}{k_\mu}~,~~~
y^\mu(s_{0})= \xi^{\nu}_{(0)} \der{\mathcal{K}^{(0)}_\nu}{q_\mu}~,~~~
u^\mu(s_{0})= \xi^{\nu}_{(0)} \der{\mathcal{K}^{(0)}_\nu}{r_\mu}~,~~~
x'^\mu(s_{1}) = -\xi^\nu_{(1)} \der{\mathcal{K}^{(1)}_\nu}{p'_\mu}~,~~~ \\
\qquad x'^\mu(s_{2}) = \xi^\nu_{(2)} \der{\mathcal{K}^{(2)}_\nu}{p'_\mu}~,~~~
y'^\mu(s_{0}) = -\xi^{\nu}_{(0)} \der{\mathcal{K}^{(0)}_\nu}{q'_\mu}~,~~~
y'^\mu(s_{3}) = \xi^\nu_{(3)} \der{\mathcal{K}^{(3)}_\nu}{ q'_\mu}~,~~~
x''^\mu(s_{2}) = -\xi^\nu_{(2)} \der{\mathcal{K}^{(2)}_\nu}{ p''_\mu}~,~~~
z'^\mu(s_{2}) = -\xi^\nu_{(2)} \der{\mathcal{K}^{(2)}_\nu}{k'_\mu}~,~~~\\
y''^\mu(s_{3}) = -\xi^\nu_{(3)} \der{\mathcal{K}^{(3)}_\nu}{ q''_\mu}~,~~~
u'^\mu(s_{3}) = -\xi^\nu_{(3)} \der{\mathcal{K}^{(3)}_\nu}{r'_\mu}~ \ .
\end{gather*}
It is easy to check that the above equations of motion and boundary conditions
are invariant under the following translation transformations:
\begin{equation}
 \begin{split}
 x^\mu_B&= x^\mu_A+ b^\nu\{(k\oplus p)_\nu,x^\mu\}\ ,\\
 z^\mu_B&= z^\mu_A+ b^\nu\{(k\oplus p)_\nu ,z^\mu\}  \ ,\\
  y^\mu_B&= y^\mu_A+ b^\nu\{(r \oplus q)_\nu ,y^\mu\} \ ,\\
 u^\mu_B&= u^\mu+ b^\nu\{(r \oplus q)_\nu ,u^\mu\}  \ ,\\
{x'}^\mu_B&={x'}_{A}^{\mu}+ b^\nu\{p'_\nu ,{x'}^\mu\}\ ,\\
{y'}_{B}^{\mu}&={y'}_{A}^{\mu}+ b^\nu\{q'_\nu,{y'}^{\mu}\}\ ,\\
{x''}^\mu_B&= x''^\mu_A+ b^\nu\{(k'\oplus p'')_\nu ,x''^\mu\} \ ,\\
{z'}^\mu_B&= z'^\mu_A+ b^\nu\{k'\oplus p'')_\nu ,z'^\mu\} \ ,\\
{y''}^\mu_B&=y''^\mu_A+ b^\nu\{(r' \oplus q'')_\nu ,y''^\mu\} \ ,\\
{u'}^\mu_B&= u'^\mu_A+ b^\nu\{(r' \oplus q'')_\nu ,u'^\mu\} \ ,
 \label{translations}
\end{split}
\end{equation}
where $b^\mu$ are the translation parameters and it is understood that
$\{z^{\mu} , k_{\nu} \} = \delta^{\mu}_{\nu}$, $\{x^{\mu} , p_{\nu} \} = \delta^{\mu}_{\nu}$,
$\{y^{\mu} , q_{\nu} \} = \delta^{\mu}_{\nu}$, $\{u^{\mu} , r_{\nu} \} = \delta^{\mu}_{\nu}$, $\{z'^{\mu} , k'_{\nu} \} = \delta^{\mu}_{\nu}$, $\{x'^{\mu} , p'_{\nu} \} = \delta^{\mu}_{\nu}$,
$\{y'^{\mu} , q'_{\nu} \} = \delta^{\mu}_{\nu}$, $\{u'^{\mu} , r'_{\nu} \} = \delta^{\mu}_{\nu}$, $\{x''^{\mu} , p''_{\nu} \} = \delta^{\mu}_{\nu}$, $\{y''^{\mu} , q''_{\nu} \} = \delta^{\mu}_{\nu}$.

Because of relative locality we evidently need here two observers Alice and Bob chosen so that the questions
here of interest can be investigated in terms of the locality of interactions near them.
As announced we focus on the case in which the interactions ${\cal K}^{(0)}$
and ${\cal K}^{(1)}$ are coincident, and we take as Alice an observer for whom these two interactions occur in the
origin of her reference frame. This in particular allows us to restrict our attention to cases
with $x'^\mu_A(s_1)=y'^\mu_A(s_0)=0$.
We take the other observer, Bob, at rest with respect to Alice and such that the event ${\cal K}^{(3)}$ occurs in the origin
of Bob's reference frame, so that $y'^\mu_B(s_3)=0$.
Since in the $\kappa$-momentum case the physical speed of ultrarelativistic particles depends on their energy \cite{anatomy}
the interaction ${\cal K}^{(2)}$ cannot be coincident with the interaction ${\cal K}^{(3)}$ (since ${\cal K}^{(0)}$ and ${\cal K}^{(3)}$
exchange a soft particle whereas ${\cal K}^{(1)}$  and ${\cal K}^{(2)}$ exchange a hard particle one must take into account
the difference in physical speed between the hard and the soft exchanged particle).
But this dependence on energy of the physical speed of ultrarelativistic particles is anyway a small $\ell$-suppressed effect,
so we can focus on a situation where ${\cal K}^{(2)}$ and ${\cal K}^{(3)}$ are nearly coincident, and we study that
situation assuming ${\cal K}^{(2)}$ occurs in spatial origin  of Bob's reference frame (but at time different from ${\cal K}^{(3)}$).
 This allows us to specify   $x'^1_B(s_2)=0$.
 Also note that as long as the distance of ${\cal K}^{(2)}$ from the spacetime origin of  Bob's reference frame is an $\ell$-suppressed
 feature Bob's description of the locality (or lack thereof) of the interaction ${\cal K}^{(2)}$ is automatically immune from
 relative-locality effects at leading order in $\ell$, which is the order at which we are working.

 Equipped with this choice of observers and these simplifying assumptions about the relevant events, we can
 quickly advance with our analysis of causal links from the relative-locality perspective.
We start by noticing that from the equations of motion it follows that both for Alice and Bob\footnote{Note that within our conventions
the direction of propagation and the sign of the spatial momentum with lower index, $p_1$, are opposite. So negative $p_1$ is actually
for propagation along the positive direction of the $x^1$-axis.}
\begin{equation}
 \frac{\dot{x}'^1}{\dot{x}'^0}=1-\ell p'_1 \, , \qquad \frac{\dot{y}'^1}{\dot{y}'^0}=1\, .
 \end{equation}
This implies that according to Alice (for whom the events ${\cal K}^{(0)}$
and ${\cal K}^{(1)}$ occur in the origin of the reference frame)
the worldlines of the two exchanged particles are
\begin{equation}
 \begin{split}
 {x'}_{A}^{1}&=(1-\ell p'_1){x'}_{A}^{0}\ ,\\
 {y'}_{A}^{1}&={y'}_{A}^{0}\ .
 \label{eq-motion-alice}
\end{split}
\end{equation}
A key aspect
of the analysis we are reporting in this section is establishing how these
two worldlines are described by the distant observer Bob.
On the basis of (\ref{translations}) one concludes that the relevant translation transformation is undeformed:
\begin{equation}\label{translations_atoms}
\begin{split}
 {x'}_{B}^{\mu}(s)&={x'}_{A}^{\mu}(s)+b^{\nu}  \{ p'_\nu , x'^\mu\} = {x'}_{A}^{\mu}(s)-b^{\mu} \ , \\
 {y'}_{B}^{\mu}(s)&={y'}_{A}^{\mu}(s)+b^{\nu}  \{ q'_\nu , y'^\mu\}= {y'}_{A}^{\mu}(s)-b^{\mu}\ .
\end{split}
 \end{equation}
 So the worldlines in Bob's coordinatization must have the form
 \begin{equation}
  \begin{split}
   x'^1_B&=(1-\ell p'_1)x'^0_B -b^1+b^0-b^0\ell p'_1\, , \\
   y'^1_B&=y'^0_B -b^1+b^0 \, .
   \label{eq-motion-bob}
  \end{split}
 \end{equation}
 Since we have specified for Bob that ${\cal K}^{(3)}$ occurs in the  origin of his reference frame,  $y'^\mu_B(s_3)=0$,
 we must have that $b^0=b^1$. And then finally we establish that the event ${\cal K}^{(2)}$, occurring in the
  spatial origin of Bob's reference frame, $x^1_B(s_2)=0$, is timed by Bob at
 \begin{equation}
   x'^0_B(s_2)=b^1\ell p'_1 \ .
   \label{arrival-time}
 \end{equation}
In particular for positive $\ell$ one has that according to  Bob  ${\cal K}^{(2)}$ occurs before ${\cal K}^{(3)}$ in his spatial origin,
with a time difference between them given by $\Delta t=b^1\ell p'_1$.

This was just preparatory material for the point we most care about in this section, which concerns possible
paradoxes for causality  and their clarification.
For that we need to look at how Alice describes the two events distant from her, ${\cal K}^{(2)}$ and ${\cal K}^{(3)}$.
${\cal K}^{(3)}$ is an interaction involving only soft particles so nothing noteworthy can arise from looking
at ${\cal K}^{(3)}$, but ${\cal K}^{(2)}$ involves hard particles and therefore the inferences about ${\cal K}^{(2)}$ by
observer Alice, who is distant from
 ${\cal K}^{(2)}$, will give a description of  ${\cal K}^{(2)}$ as an apparently non-local interaction.
This is the main implication of relative locality, and we can see that it does give rise to a
combined description of ${\cal K}^{(2)}$ and ${\cal K}^{(3)}$ that at first may appear puzzling from the causality perspective.
We show this by noting down the values of coordinates of particles involved in ${\cal K}^{(2)}$ and ${\cal K}^{(3)}$ according to
Alice. For the particles with coordinates $y''^{\mu}$ and $u'^\mu$
on the basis of (\ref{translations}) one finds that the translation is completely undeformed, and since
$y''^{\mu}_B (s_3)=u'^\mu_B (s_3)=0$ one has that
$$y''^{\mu}_A (s_3)=u'^\mu_A (s_3)=\xi_{(3)A}^{\mu}=b^1 \ .$$
For the particles involved in the hard vertex ${\cal K}^{(2)}$, with coordinates $x''^\mu$
and $z'^\mu$, on the basis of (\ref{translations})
one finds that the translation is deformed, and starting from the fact
that $x''^0_B=b^1\ell p'_1$, $x''^1_B=0$, $z'^0_B=b^1\ell p'_1$, $z'^1_B=0$
one arrives at finding that
\begin{equation}
\begin{split}
  x''^0_A (s_2)&=b^1+b^1\ell p'_1 \ , \\
  x''^1_A (s_2)&=b^1 - b^1 \ell k'_1\ \approx b^1, \\
  z'^0_A (s_2)&= b^1+b^1\ell p'_1 -b^1\ell p''_1\ , \\
  z'^1_A (s_2)&= b^1 \ .
\end{split}
\end{equation}
As shown in Fig.4 the most striking situation from the viewpoint of causality arises
when  $p'_1 \simeq  p''_1$, in which case according to Alice $z'^0_A (s_2)= z'^1_A (s_2) =b^1$,
which means that the particle with coordinates $z'^\mu$, who actually interacts at ${\cal K}^{(2)}$,
in the coordinatization by distant observer Alice appears to come out of the interaction ${\cal K}^{(3)}$.
This is an example of the sort of apparent paradoxes for causality that can be encountered with relative locality:
they all concern the description of events by distant observers.
Of course, there is no true paradox
since a known consequence of relative locality is that inferences about distant events are misleading.
Indeed, as also shown in Fig.4, Bob's description of the interactions
${\cal K}^{(2)}$ and ${\cal K}^{(3)}$ (which are near Bob) is completely unproblematic.
However, in turn, Bob's inferences about the events ${\cal K}^{(0)}$ and ${\cal K}^{(1)}$
(which are distant from Bob) are affected by peculiar relative-locality features, as also shown in Fig.4. In looking at Fig.4 readers should also keep in mind that for that figure we magnified effects in order to render them visible: actually all noteworthy features in Fig.4 are Planck-scale suppressed, and would amount to time intervals no greater than $10^{-19}s$ for Earth experiments (over distances of, say, $10^6m$) involving particles with currently accessible energies (no greater than, say, $1TeV$).

\newpage

\begin{figure}[h!]
\includegraphics[width= 0.5\columnwidth]{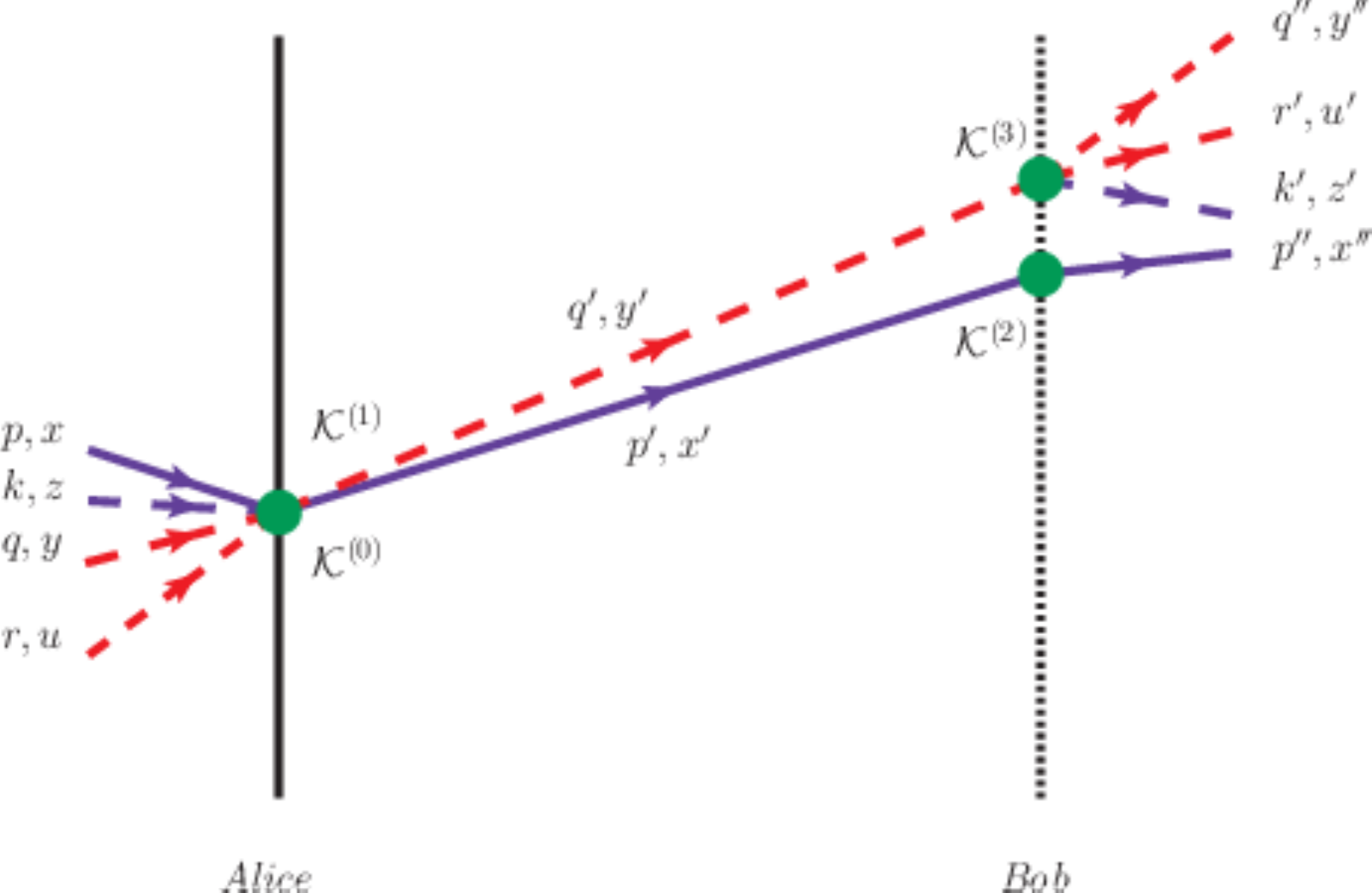}
\includegraphics[width= 0.5\columnwidth]{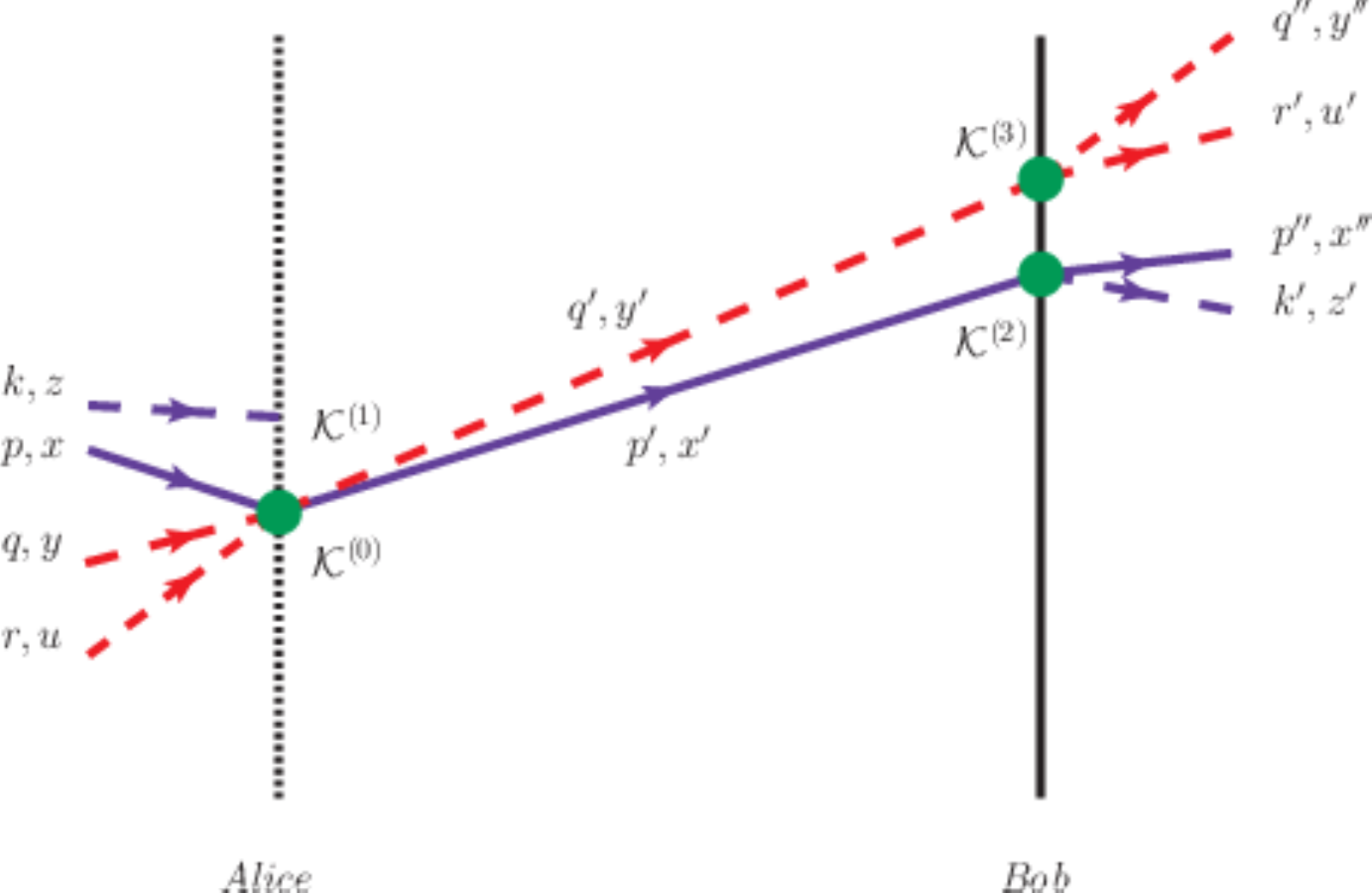}
\caption{{\small The two causally-connected pairs of events considered in this section can lead to a striking picture of distant inferences (because of relative locality)\ when  $p'_1 \simeq  p''_1$.
In that case the particle with
coordinates $z'^\mu$, who actually interacts at ${\cal K}^{(2)}$,
in the coordinatization by distant observer Alice (top panel)\ appears to come out of the interaction ${\cal K}^{(3)}$.
In turn, as we  show in the bottom panel of the figure, Bob's inferences about the events ${\cal K}^{(0)}$ and ${\cal K}^{(1)}$
(which are distant from Bob) are affected by peculiar relative-locality features.}}
\end{figure}

\section{Causal Loops}
The observations on relative locality reported in the previous two sections illustrate how misleading the characterization of events and chains of
events can be, if not based on how each event is seen by a nearby observer.
For chains of events this imposes that the analysis be based on more than one observer: at least one observer for each
interaction in the chain.

Equipped with this understanding we now progress to the next level in testing causality: we consider the possibility
of a ``causal loop", {\it i.e.} a
chain of events that form a loop in such a way that causality would
be violated.

The starting point for being concerned about these causal loops is the analysis reported in Ref.\cite{linq}, which considered
a loop diagram of the type here shown in Fig.5. Ref.\cite{linq} works on a curved momentum space, but without
enforcing relative locality, and finds that a causal loop of the type here shown in Fig.5 could be possible.
Our objective is to show that such causal loops are excluded
if one enforces relative locality. In light of the observations reported in the previous two sections we shall of course need to study the loop diagram
in Fig.5 on the basis of the findings of two observers, one near the first interaction and one near the second interaction
(whereas the analysis of Ref.\cite{linq} only considered the perspective of one observer, in which case the principle of relativity of spacetime
locality cannot be enforced or investigated).

\begin{figure}[h!]
\begin{center}
\includegraphics[scale=0.6]{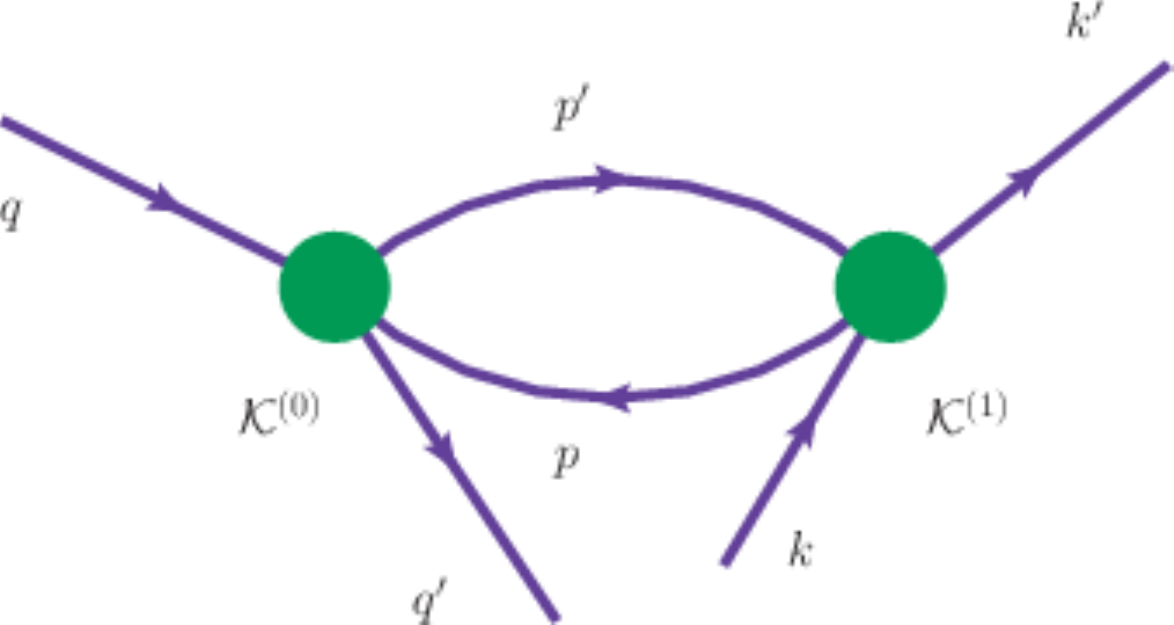}
\end{center}
\caption{\small We here show schematically a pair of events causally connected by the exchange of two particles arranged in such a way that one would have a causal loop. Such causal loops are allowed, if one assumes curvature of momentum space without enforcing (DSR-)relativistic covariance and the associated relativity of spacetime locality.}
\label{looplinq}
\end{figure}

We stress that here, just as in Ref.\cite{linq}, we are working at the level of classical mechanics,
so the loop diagram in Fig.\ref{looplinq} involves all particles on-shell and merely keeps track of the causal links among different events, assigning worldlines exiting/entering each event (one should not confuse such loop diagrams
with the different notion arising in Feynman's perturbative approach to quantum field theory).

We start by writing down an action of the type already considered in the previous two sections,
which gives the description of the loop diagram in Fig.\ref{looplinq} within the relative-locality curved-momentum-space
formalism proposed in Ref.\cite{principle}.
We shall see that our action does reproduce the equations of motion and the boundary conditions
which were at the basis of the analysis reported in Ref.\cite{linq}. This action giving the diagram in Fig. \ref{looplinq} is
\begin{equation}
\begin{array}{lll}
\mathcal S      &=&     \displaystyle
\int_{-\infty}^{s_0}ds\left(y^\mu\dot q_\mu+\mathcal N_q(\mathcal C_q-m^2_q)\right)
+\int_{s_0}^{+\infty}ds\left(y'^\mu\dot q_\mu'+\mathcal N_{q'}(\mathcal C_{q'}-m^2_{q'})\right)
+\int_{-\infty}^{s_1}ds\left(z^\mu\dot k_\mu+\mathcal N_k(\mathcal C_k-m^2_k)\right)+\\

&&\displaystyle
+\int_{s_1}^{+\infty}ds\left(z'^\mu\dot k'_\mu+\mathcal N_{k'}(\mathcal C_{k'}-m^2_{k'})\right)
+\int_{s_0}^{s_1}ds\left(x'^\mu\dot p'_\mu+\mathcal N_{p'}(\mathcal C_{p'}-m^2_{p'})\right)
+\int_{s_1}^{s_0}ds\left(x^\mu\dot p_\mu+\mathcal N_p(\mathcal C_p-m^2_p)\right)+\\

&&\displaystyle
-\xi_{(0)}^{\mu}\mathcal K_{\mu}^{(0)}-\xi_{(1)}^{\mu}\mathcal K_{\mu}^{(1)} \ ,
\end{array}
\end{equation}
where $\mathcal K_\mu^{(0)}=\left[(q\oplus p)\oplus(\ominus(p'\oplus q'))\right]_\mu$ and
$\mathcal K_\mu^{(1)}=\left[(p'\oplus k)\oplus(\ominus(k'\oplus p))\right]_\mu$. It is important for us to stress, since this is the key ingredient
for seeking a violation of causality, that the last integral,
which stands for the free propagation of the particle which is travelling back in time,
has inverted integration extrema.

By varying this action we obtain equations of motion
\begin{subequations}
\begin{gather}
\dot p_\mu=0 \ , \qquad \dot p'_\mu=0 \ , \qquad \dot q_\mu=0 \ , \qquad \dot q'_\mu=0 \ ,\qquad \dot k_\mu=0 \ , \qquad \dot k'_\mu=0 \ ,\\
\mathcal C_p=m^2_p \ , \qquad \mathcal C_{p'}=m^2_{p'} \ , \qquad \mathcal C_q=m^2_q \ , \qquad \mathcal C_{q'}=m^2_{q'} \ , \qquad \mathcal C_{k'}=m^2_{k'} \ , \qquad \mathcal C_k=m^2_k \ ,\\
\dot x^\mu(s)=\mathcal N_p\der{\mathcal C_p}{p_\mu} \ ,\label{dotcoordinates1}\qquad
\dot x'^\mu(s)=\mathcal N_{p'}\der{\mathcal C_{p'}}{p'_{\mu}} \ ,\qquad
\dot y^\mu(s)=\mathcal N_q\der{\mathcal C_q}{q_\mu} \ ,\\
\dot y'^\mu(s)=\mathcal N_{q'}\der{\mathcal C_{q'}}{q'_\mu} \ ,\label{dotcoordinates2}\qquad
\dot z^\mu(s)=\mathcal N_k\der{\mathcal C_k}{k_\mu} \ ,\qquad
\dot z'^\mu(s)=\mathcal N_{k'}\der{\mathcal C_{k'}}{k'_\mu} \ , \\
\mathcal K_\mu^{(0)}=0 \ , \qquad \mathcal K_\mu^{(1)}=0 \ ,\label{boundaries1}
\end{gather}
\end{subequations}
and boundary terms
\begin{subequations}
\begin{gather}
y^\mu(s_0)=\xi_{(0)}^\nu\der{\mathcal K_\nu^{(0)}}{q_\mu} \ ,\qquad
y'^\mu(s_0)=-\xi_{(0)}^\nu\der{\mathcal K_\nu^{(0)}}{q'_{\mu}} \ ,\qquad
z'^\mu(s_1)=-\xi_{(1)}^\nu\der{\mathcal K_\nu^{(1)}}{k'_\mu} \ ,\qquad
z^\mu(s_1)=\xi_{(1)}^\nu\der{\mathcal K_\nu^{(1)}}{k_\mu} \ ,\label{boundaries2}\\
x^\mu(s_0)=\xi_{(0)}^\nu\der{\mathcal K_\nu^{(0)}}{p_\mu} \ ,\qquad
x^\mu(s_1)=-\xi_{(1)}^\nu\der{\mathcal K_\nu^{(1)}}{p_\mu} \ ,\qquad
x'^\mu(s_0)=-\xi_{(0)}^\nu\der{\mathcal K_\nu^{(0)}}{p'_\mu} \ ,\qquad
x'^\mu(s_1)=\xi_{(1)}^\nu\der{\mathcal K_\nu^{(1)}}{p'_\mu} \ ,\label{boundaries3}
\end{gather}
\end{subequations}
which indeed reproduce the ones used in the analysis reported in Ref.\cite{linq}.

\subsection{Aside on the absence of causal loops in Special Relativity}
We find it useful to start by first considering the $\ell \rightarrow 0$ limit of the problem of interest
in this section: the causal loop in Special Relativity ({\it i.e.} with Minkowskian geometry of momentum space).
This allows us to assume temporarily that the on-shellness is governed by $\mathcal C^{(0)}=p_0^2-p_1^2$
and that therefore the following relationship holds
\begin{equation}\label{minkowskianeom}
\dot x^\mu(s)=\left(\dot x^\nu\dot x_\nu\right)^\frac{1}{2}\frac{p^{\mu}}{m_p} \ .
\end{equation}
We take advantage of some simplification of analysis,  without loosing any of the conceptual ingredients here of interest, by focusing on  $\dot x_\mu\dot x^\mu>0,\ \dot x^0>0;\ p^2=m_p^2>0,\ p^0\geq m_p>0$, {\it i.e.}
our particles travel along timelike worldlines.
 We  have
that the proper time of a particle is given by
\begin{equation}
d\tau=\left(\dot x^\mu\dot x_\mu\right)^{\frac{1}{2}}ds=
\dot x^0\sqrt{1-\left(\frac{\dot x^1}{\dot x^0}\right)^{2}}ds=
\dot x^0\sqrt{1-\left(\frac{p^1}{p^0}\right)^2}ds=
\dot x^0\gamma_p^{-1}ds \ ,
\end{equation}
where $\gamma_p$ is the usual Lorentz factor and in the third equality we used  (\ref{minkowskianeom}).

Going back to the diagram in Fig.\ref{looplinq} we have that for the particle with phase-space coordinates $(p',x')$, whose worldline
 is exchanged between the interaction $\mathcal K^{(0)}$ and the interaction $\mathcal K^{(1)}$ (and therefore travels from $x'^\mu(s_0)$ to $x'^\mu(s_1)$) the following chain of equalities holds
\begin{equation}\label{fin-iniz(x')}
\begin{array}{lll}
x'^\mu(s_1)-x'^\mu(s_0)
&=&\displaystyle
\int_{s_0}^{s_1}ds\,\frac{dx'^\mu}{ds}=\int_{s_0}^{s_1}ds\left(\dot x'^\nu\dot x'_\nu\right)^\frac{1}{2}\frac{p'^\mu}{m_{p'}}=\\
&=&\displaystyle
\int_{\tau'(s_0)}^{\tau'(s_1)}d\tau'\frac{p'^\mu}{m_{p'}}=\Delta\tau'u'^\mu \ ,
\end{array}
\end{equation}
with $\displaystyle u'^\mu=\frac{p'^\mu}{m_{p'}}$ \ .

Similarly, for the other particle exchanged between  $\mathcal K^{(0)}$ and $\mathcal K^{(1)}$, the one with phase-space coordinates $(p,x)$,
one has that
\begin{equation}\label{fin-iniz(x)}
\begin{array}{lll}
x^\mu(s_0)-x^\mu(s_1)
&=&\displaystyle
\int_{s_1}^{s_0}ds\,\frac{dx^\mu}{ds}=\int_{s_1}^{s_0}ds\left(\dot x^\nu\dot x_\nu\right)^\frac{1}{2}\frac{p^\mu}{m_p}=\\
&=&\displaystyle
\int_{\tau(s_1)}^{\tau(s_0)}d\tau\frac{p^\mu}{m_p}=\Delta\tau\ u^\mu \ .
\end{array}
\end{equation}
Since in this subsection we are working
in the $\ell \rightarrow 0$ limit we have that $\mathcal K_\mu^{(0)}=q_\mu+p_\mu-p'_\mu-q'_\mu$ and $\mathcal K_\mu^{(1)}=p'_\mu+k_{\mu}-k'_\mu-p_\mu$ ,
in which case it is easy to see that our boundary conditions simply enforce
\begin{equation}
\xi_{(0)}^\mu=x'^\mu(s_0)~,~~~
\xi_{(0)}^\mu=x^\mu(s_0)~,~~~
\xi_{(1)}^\mu=x'^\mu(s_1)~,~~~
\xi_{(1)}^\mu=x^\mu(s_1)~.
\end{equation}
So evidently
\begin{gather}
\xi_{(1)}^\mu-\xi_{(0)}^\mu=x'^\mu(s_1)-x'^\mu(s_0)=\Delta\tau'u'^\mu \ ,\\
\xi_{(0)}^\mu-\xi_{(1)}^\mu=x^\mu(s_0)-x^\mu(s_1)=\Delta\tau\ u^\mu \ ,
\end{gather}
and
\begin{equation}\label{speesrelationSpRel}
\Delta\tau\ u^\mu+\Delta\tau'u'^\mu=0 \ .
\end{equation}
Since the relevant proper-time intervals are positive and the zero components of the four-velocities are positive
 this requirement can never be satisfied: as well known causal loops are forbidden in Special Relativity.

 Another way to see that causal loops are forbidden in Special Relativity can be based on deriving the relationship between the relevant proper-time intervals
  and the interaction coordinates $\xi_{(0)}^\mu$, $\xi_{(1)}^\mu$. One easily finds that
\begin{gather}
\Delta\tau=\int_{s_1}^{s_0}ds\ \dot x^0\gamma_p^{-1}=
\gamma_p^{-1}\left(x^0(s_0)-x^0(s_1)\right)=
\gamma_p^{-1}\left(\xi_{(0)}^0-\xi_{(1)}^0\right) \ ,\label{deltatauSR}\\
\Delta\tau'=\int_{s_0}^{s_1}ds\ \dot x'^0\gamma_{p'}^{-1}=
\gamma_{p'}^{-1}\left(x'^0(s_1)-x'^0(s_0)\right)=
\gamma_{p'}^{-1}\left(\xi_{(1)}^0-\xi_{(0)}^0\right) \ .\label{deltatau'SR}
\end{gather}
So again the fact that  $\Delta\tau \geq 0$ and $\Delta\tau'\geq 0$ excludes the causal loop,
since on the basis of (\ref{deltatauSR})-(\ref{deltatau'SR}) this would require $\xi_{(0)}=\xi_{(1)}$: by construction $\left(\xi_{(1)}-\xi_{(0)}\right)_\mu \left(\xi_{(1)}-\xi_{(0)}\right)^\mu \geq 0$
(the interval between the two interactions is timelike or null)
 and then $\xi_{(0)}^0=\xi_{(1)}^0$ implies  $\xi_{(0)}^\mu=\xi_{(1)}^\mu$, \textit{i.e.} the loop can only collapse
 into a single event (no causality issue, not a causal loop).

\subsection{Causal loop with curved momentum space}
Our next step is to introduce leading-order-in-$\ell$ corrections, but without enforcing the principle
of relative locality. Such setups in general do allow causal loops, as we shall now show (in agreement with what was already claimed in Ref.\cite{linq}).
What changes with respect to the special-relativistic case of the previous subsection
is that (for the $\kappa$-momentum case, which we chose as illustrative example)
the on-shellness is governed by $\mathcal C_p=p_0^2-p_1^2-\ell p_0p_1^2$
while conservation laws at first order take the form
\begin{subequations}
\begin{align}
&\mathcal K_    {\ 0}^{(0)}=q_0+p_0-q'_0-p'_0 \ ,\\
&\mathcal K_{\ 1}^{(0)}=q_1+p_1-q'_1-p'_1-\ell\left[q_0p_1-(q_0+p_0-q'_0-p'_0)p'_1-(q_0+p_0-q'_0)q'_1\right] \ ,\\
&\mathcal K_    {\ 0}^{(1)}=p'_0+k_0-p_0-k'_0 \ ,\\
&\mathcal K_{\ 1}^{(1)}=p'_1+k_1-p_1-k'_1-\ell\left[p'_0k_1-(p'_0+k_0-p_0-k'_0)k'_1-(p'_0+k_0-p_0)p_1\right] \ .
\end{align}
\end{subequations}
Also the equations of motion are $\ell$-deformed, as shown in (\ref{dotcoordinates1})-(\ref{dotcoordinates2}),
and for example one has that
\begin{equation}\label{kminkowskianeom}
\dot x^\mu(s)=\mathcal N_p\left[2p^\mu -\ell\left(\delta^\mu_0p_1^2+\delta^\mu_12p_0p_1\right)\right] \ .
\end{equation}
This still allows one to write a relationship analogous to (\ref{minkowskianeom}) from the previous subsection,
\begin{equation}
\dot x^\mu(s)=\left(\dot x^\nu\dot x_\nu\right)^\frac{1}{2}u^\mu \ ,
\end{equation}
but with
 $$u^\mu=\frac{p^\mu}{m_p}-\frac{\ell}{2m_p}\left(-2p^\mu\frac{p_0p_1^2}{m_p^2}+\delta_0^\mu p_1^2+\delta^\mu_1 2p_0p_1\right)\ .$$
Analogously, for $x'^\mu$ one has that
\begin{equation}
\dot x'^\mu(s)=\left(\dot x'^\nu\dot x'_\nu\right)^\frac{1}{2}u'^\mu \ ,
\end{equation}
with
 $$u'^\mu=\frac{p'^\mu}{m_{p'}}-\frac{\ell}{2m_{p'}}\left(-2p'^\mu\frac{p'_0p_1'^2}{m_{p'}^2}+\delta_0^\mu p_1'^2+\delta^\mu_1 2p'_0p'_1\right)\ .$$

In close analogy with (\ref{fin-iniz(x')}) and (\ref{fin-iniz(x)}) one easily finds that
\begin{align}
&x'^\mu(s_1)-x'^\mu(s_0)=\Delta\tau'u'^\mu, \label{kfin-iniz(x')}\\
&x^\mu(s_0)-x^\mu(s_1)=\Delta\tau\ u^\mu ~, \label{kfin-iniz(x)}
\end{align}
and from (\ref{boundaries3}) it follows that
\begin{align}
&\xi_{(0)}^\nu=-x'^\mu(s_0)\left(\der{\mathcal K_\nu^{(0)}}{p'_\mu}\right)^{-1}=x^\mu(s_0)\left(\der{\mathcal K_\nu^{(0)}}{p_\mu}\right)^{-1} \ ,\label{boundaries3new}\\
&\xi_{(1)}^\nu=x'^\mu(s_1)\left(\der{\mathcal K_\nu^{(1)}}{p'_\mu}\right)^{-1}=-x^\mu(s_1)\left(\der{\mathcal K_\nu^{(1)}}{p_\mu}\right)^{-1} \ .\label{boundaries4new}
\end{align}
Combining (\ref{boundaries3new}) with (\ref{kfin-iniz(x)}) one finds that
\begin{equation}\label{eq:-3}
-x'^\mu(s_0)\left(\der{\mathcal K_\nu^{(0)}}{p'_\mu}\right)^{-1}\der{\mathcal K_\nu^{(0)}}{p_\rho}=x^\rho(s_0)=x^\rho(s_1)+\Delta\tau u_p^\rho \ ,
\end{equation}
while
combining (\ref{boundaries4new}) with (\ref{kfin-iniz(x')})
one finds that
\begin{equation}\label{eq:-4}
x^\rho(s_1)=
-x'^\mu(s_1)\left(\der{\mathcal K_\nu^{(1)}}{p'_\mu}\right)^{-1}\der{\mathcal K_\nu^{(1)}}{p_\rho}=
-(x'^\mu(s_0)+\Delta\tau'u'^\mu)\left(\der{\mathcal K_\nu^{(1)}}{p'_\mu}\right)^{-1}\der{\mathcal K_\nu^{(1)}}{p_\rho} \ .
\end{equation}
Finally, combining (\ref{eq:-4}) with (\ref{eq:-3}), we obtain the same condition given in \cite{linq},
\begin{equation}\label{eq:-5}
\left[\der{\mathcal K_\nu^{(1)}}{p_\rho}\left(\der{\mathcal K_\nu^{(1)}}{p'_\mu}\right)^{-1}-\der{\mathcal K_\nu^{(0)}}{p_\rho}\left(\der{\mathcal K_\nu^{(0)}}{p'_\mu}\right)^{-1}\right]x'^\mu(s_0)=
-\der{\mathcal K _\nu^{(1)}}{p_\rho}\left(\der{\mathcal K_\nu^{(1)}}{p'_\mu}\right)^{-1}\Delta\tau'u'^\mu+\Delta\tau u^\rho \ ,
\end{equation}
which takes the following form upon expanding $\mathcal K_\nu^{(0)}$
and $\mathcal K_\nu^{(1)}$
 to leading order in $\ell$:
\begin{equation}\label{eq:-5'}
\ell\left[\delta_1^\rho\left(k'_0-q_0\right)+\delta_0^\rho\left(q'_1-k_1\right)\right]x'^1(s_0)=
\Delta\tau u^\rho+\Delta\tau'\left[u'^\rho+u'^1\ell\left(\delta_0^\rho k_1-\delta_1^\rho k'_0\right)\right] \ .
\end{equation}
This (\ref{eq:-5'}) is what replaces (\ref{speesrelationSpRel}) when the causal loop is analyzed on a curved momentum
space without enforcing relative locality.

Notice that this (\ref{eq:-5'}), when its left-hand side does not vanish,  can have solutions
with positive $\Delta\tau$ and $\Delta\tau'$ and positive zero components of the four-velocities,
which was not possible with (\ref{speesrelationSpRel}).
This means that contrary to the special-relativistic case (Minkowski momentum space)
causal loops are possible on a curved momentum
space, at least if one does not enforce relative locality.

We also note down some equalities that follow
from  (\ref{eq:-5'})
 and therefore must hold for the causal loop to be allowed
 \begin{equation}\label{sistemalinqcompletoA}
\Delta\tau=-\Delta\tau'\frac{u'^0}{u^0}+\ell x'^1(s_0)\left(\frac{q'_1-k_1}{u^0}\right)-\ell\Delta\tau'\left(\frac{u'^1k_1}{u^0}\right) \ ,
\end{equation}
\begin{equation}\label{sistemalinqcompletoB}
\ell x'^1(s_0)=\Delta\tau'\frac{u^1u'^0-u^0u'^1+\ell u'^1(k_1u^1+k'_0u^0)}{u^0(q_0-k'_0)+u^1(q'_1-k_1)}
\end{equation}
and we note that in order for (\ref{sistemalinqcompletoA})
to have acceptable solutions one must have that
\begin{equation}
x'^1(s_0)>\frac{\Delta\tau' (u'^0+\ell u'^1k_1)}{\ell|q'_1-k_1|} \ .
\label{reqextra}
\end{equation}
This is in good agreement with the results of Ref.~\cite{linq}, but we find
useful to add some observations to those reported in Ref.~\cite{linq}.
A first point to notice is that Eq.~(\ref{reqextra}) appears to suggest
that  $x'^1$ should take peculiarly large values, as in some of the
estimates given in Ref.~\cite{linq}, since $x'^1$ has magnitude set by
a formula with the small scale $\ell$ in the denominator.
If one could conclude that only cases with ultralarge $x'^1$ allowed
such a causal loop, then the violations of causality would be to some extent
less concerning (if confined to a range of values of $x'^1$ large enough to
fall outside our observational window). However, it is easy to see
that (\ref{reqextra}) does not really impose any restriction on the
size of $x'^1$: one will have that typically $x'^1$ is much larger than
$\Delta\tau'$ but there are causal loops for any value of $x'^1$ (under the condition
of taking suitable values of $\Delta\tau'$  and $\Delta\tau$).
So when momentum space is curved and one does not enforce
the relativity of spacetime locality the violations
of causality are rather pervasive.

There is also a technical point that deserves some comments
and is related to this pervasiveness
of the violations of causality: it might appear to be surprising that
within a perturbative expansion, assuming small $\ell$, one
arrives at a formula like (\ref{reqextra}), with $\ell$ in the denominator.
This is however not so surprising considering the role of $x'^1$
in this sort of analysis. The main clarification comes from observing
that in the unperturbed theory (the $\ell=0$ theory, {\it i.e.}
special relativity) $x'^1$ is completely undetermined: as shown in the
previous subsection the only causal loops allowed in special relativity
are those that collapse (no violation of causality) and such collapsed
causal loops are allowed for any however large or however small value
of $x'^1$. As stressed above this fact that $x'^1$ can take any value
is preserved by the $\ell$ corrections. The apparently surprising factor
of $1/\ell$ only appears in a relationship between  $x'^1$ and $\Delta\tau'$.
If $x'^1$ and $\Delta\tau'$ both had some fixed finite value in the $\ell=0$ theory
than at finite small $\ell$ their values should change by very little.
But since in the $\ell=0$ theory $x'^1$ is unconstrained  (in particular
it could take any however large value)  and its value is
not linked in any way to the value $\Delta\tau'$, then it is
not surprising that the $\ell$ corrections take the form shown
for example in (\ref{reqextra}).

\subsection{Causal loop analysis in 3+1 dimensions}

So far we examined the 1+1-dimensional case, but it is rather evident that the features we discussed in the previous subsection are not an artifact of that dimensional reduction. Nonetheless it is worth pausing briefly in this subsection for verifying that indeed those features are still present in $3+1$ dimensions.
In this case the on-shellness is governed by $\mathcal C_p=p_0^2-\vec{p}^2-\ell p_0\vec{p}^2$
while conservation laws at first order take the form
\begin{subequations}
\begin{align}
&\mathcal K_    {\ 0}^{(0)}=q_0+p_0-q'_0-p'_0 \ ,\\
&\mathcal K_{\ i}^{(0)}=q_i+p_i-q'_i-p'_i-\ell\delta_i^j\left[q_0p_j-(q_0+p_0-q'_0-p'_0)p'_j-(q_0+p_0-q'_0)q'_j\right] \ ,\\
&\mathcal K_    {\ 0}^{(1)}=p'_0+k_0-p_0-k'_0 \ ,\\
&\mathcal K_{\ i}^{(1)}=p'_i+k_i-p_i-k'_i-\ell\delta_i^j\left[p'_0k_j-(p'_0+k_0-p_0-k'_0)k'_j-(p'_0+k_0-p_0)p_j\right] \ ,
\end{align}
\end{subequations}
where $i,j=1,2,3$.

Adopting these expressions, eq.(\ref{eq:-5}), keeping only terms up to first order in $\ell$ in the matrices like $\der{\mathcal K_\nu^{(0)}}{p_\rho}$ and their products, takes the form
\begin{equation}\label{eq:-5''}
\ell\left[\delta_i^\rho\left(k'_0-q_0\right)+\delta_0^\rho\left(q'_i-k_i\right)\right]x'^i(s_0)=\left[u'^\rho+u'^i\ell\left(\delta_0^\rho k_i-\delta_i^\rho k'_0\right)\right]\Delta\tau'+
u^\rho\Delta\tau  \ ,
\end{equation}
or, more clearly, using the energy conservation laws,
\begin{equation}\label{4Dloop}
\begin{split}
\ell (q'_1-k_1)x'^1(s_0)+\ell (q'_2-k_2)x'^2(s_0)+\ell (q'_3-k_3)x'^3(s_0)&=(u'^0+\ell k_1 u'^1+\ell k_2 u'^2+\ell k_3 u'^3)\Delta \tau'+u^0\Delta \tau,\\
\ell(k_0-q'_0)x'^1(s_0)&=(1-\ell k'_0)u'^1\Delta\tau'+u^1\Delta\tau,\\
\ell(k_0-q'_0)x'^2(s_0)&=(1-\ell k'_0)u'^2\Delta\tau'+u^2\Delta\tau,\\
\ell(k_0-q'_0)x'^3(s_0)&=(1-\ell k'_0)u'^3\Delta\tau'+u^3\Delta\tau.
\end{split}
\end{equation}

Without really loosing any generality we can analyze the implications of this for an observer orienting her axis of the reference frame  so that $p_i=0$ and $p'_i=0$ for $i=2,3$. As a result we also
have that $u^i=0$ and $u'^i=0$ for $i=2,3$. For what concerns the other momenta
involved in the analysis, $q,\,q',\,k,\,k'$.
this choice of orientation of axis only affects rather mildly the conservation laws:
\begin{gather*}
q_2=q'_2-\ell p'_0q'_2,\qquad q_3=q'_3-\ell p'_0q'_3,\qquad q'_2=q_2+\ell p'_0q_2,\qquad q'_3=q_3+\ell p'_0q_3,\\
k_2=k'_2+\ell p'_0k'_2,\qquad k_3=k'_3+\ell p'_0k'_3,\qquad k'_2=k_2-\ell p'_0k_2,\qquad k'_3=k_3-\ell p'_0k_3.
\end{gather*}
Since $u^i=0$ and $u'^i=0$ for $i=2,3$ the last two equations of eq.(\ref{4Dloop}) imply $x'^2=0$ and $x'^3=0$, which in turn (looking then at the first two equations of  eq.(\ref{4Dloop}))
take us back to (\ref{sistemalinqcompletoA})-(\ref{sistemalinqcompletoB})
\begin{equation*}\label{sistemalinqcompletoA2}
\Delta\tau=-\Delta\tau'\frac{u'^0}{u^0}+\ell x'^1(s_0)\left(\frac{q'_1-k_1}{u^0}\right)-\ell\Delta\tau'\left(\frac{u'^1k_1}{u^0}\right),
\end{equation*}
\begin{equation*}\label{sistemalinqcompletoB2}
\ell x'^1(s_0)=\Delta\tau'\frac{u^1u'^0-u^0u'^1+\ell u'^1(k_1u^1+k'_0u^0)}{u^0(q_0-k'_0)+u^1(q'_1-k_1)} \ .
\end{equation*}
Evidently then all the features discussed for the 1+1-dimensional
in the previous subsection are also present in the 3+1-dimensional case.

\subsection{Enforcing Relative Locality}
We shall now show that our causal loop is not allowed in theories with curved momentum space if one makes sure that these theories are (DSR-)relativistic, with associated  relative locality. This suggests that relative locality is evidently a weaker notion than absolute locality but it is still strong enough to enforce causality.

By definition \cite{principle} relative locality is such that the locality of events may not be manifest in coordinatizations by distant observers, but for the coordinatizations by observers near an event (ideally at the event)  it enforces locality in a way that is \underline{no weaker} than ordinary locality.

Also notice that the definition of relative locality \underline{imposes} that translation transformations be formalized in the theory: since one must verify that events are local according to nearby observers (while they may be described as non-local by distant observers)  one must use translation transformations in order to ensure that the principle of relative locality \cite{principle} is enforced. Since our interest is in (DSR-)relativistic theories, of course
such translation transformations must be symmetries.

In Ref. \cite{anatomy} some of us introduced a prescription for having a very powerful
implementation of translational invariance in relative-locality theories. One can easily see that the causal loop described in the previous subsections is not compatible with that strong implementation of translational invariance. Evidently then we have that causality is preserved in theories with curved momentum spaces if the strong notion of translational invariance of  Ref. \cite{anatomy} is enforced by postulate.

What we here want to show is that the causal loop of Fig.5 is still forbidden even without enforcing such a strong notion of translational invariance. Causal loops are forbidden even by a minimal notion of translational invariance, the bare minimum needed in order to contemplate relative locality with a DSR-relativistic picture.

Consistently with this objective we ask only for the availability of some translation generator (with possibly complicated  momentum dependence) that can enforce the covariance of the equations of motion and the boundary conditions. Let us call our first observer Alice and the second one Bob, purely translated by a parameter $b^\mu$ with respect to Alice. For the particles involved inside the loop we have
\begin{gather}
x_B^\mu(s)=x_A^\mu(s)-b^\nu\mathcal T_\nu^\mu \ ,\label{weaker_translation_x} \\
x_B'^\mu(s)=x_A'^\mu(s)-b^\nu\mathcal T'^\mu_\nu \ ,\label{weaker_translation_x'}
\end{gather}
where $\mathcal T_\nu^\mu$ and $\mathcal T_\nu'^\mu$ are to be determined through the request of
translational invariance.

Combining the first two  boundary conditions of (\ref{boundaries3})
with  (\ref{weaker_translation_x}) we obtain
\begin{gather}
-\xi_{B(1)}^\nu\der{\mathcal K_\nu^{(1)}}{p_\mu}=
x_B^\mu(s_1)=
x_A^\mu(s_1)-b^\nu\mathcal T_\nu^\mu=
-\xi_{A(1)}^\nu\der{\mathcal K_\nu^{(1)}}{p_\mu}-b^\nu\mathcal T_\nu^\mu \label{trans_vs_boundaries_x1} \ ,\\
\xi_{B(0)}^\nu\der{\mathcal K_\nu^{(0)}}{p_\mu}=
x_B^\mu(s_0)=
x_A^\mu(s_0)-b^\nu\mathcal T_\nu^\mu=
\xi_{A(0)}^\nu\der{\mathcal K_\nu^{(0)}}{p_\mu}-b^\nu\mathcal T_\nu^\mu ~. \label{trans_vs_boundaries_x0}
\end{gather}
We find convenient to introduce $\delta\xi_{(i)}^\nu \equiv \xi_{B(i)}^\nu-\xi_{A(i)}^\nu$
and to rewrite equations (\ref{trans_vs_boundaries_x1}) and (\ref{trans_vs_boundaries_x0})
as follows
\begin{gather}
b^\nu\mathcal T_\nu^\mu=\delta\xi_{(1)}^\nu\der{\mathcal K_\nu^{(1)}}{p_\mu} \ ,\\
b^\nu\mathcal T_\nu^\mu=-\delta\xi_{(0)}^\nu\der{\mathcal K_\nu^{(0)}}{p_\mu} \ .
\end{gather}
This shows that any form one might speculate about for what concerns translational
invariance will still inevitably require enforcing
\begin{equation}\label{eq:}
\delta\xi_{(1)}^\nu\der{\mathcal K_\nu^{(1)}}{p_\mu}=-\delta\xi_{(0)}^\nu\der{\mathcal K_\nu^{(0)}}{p_\mu} \ .
\end{equation}
Similarly, combining the last two boundary conditions of (\ref{boundaries3}) with the transformation (\ref{weaker_translation_x'}) we obtain
\begin{gather}
-\xi_{B(0)}^{\nu}\der{\mathcal K_\nu^{(0)}}{p'_\mu}=
x_{B}'^\mu(s_{0})=
x_A'^\mu(s_0)-b^\nu\mathcal T_\nu'^\mu=
-\xi_{A(0)}^\nu\der{\mathcal K_\nu^{(0)}}{p'_\mu}-b^\nu\mathcal T_\nu'^\mu \ ,\\
\xi_{B(1)}^\nu\der{\mathcal K_\nu^{(1)}}{p'_\mu}=
x_B'^\mu(s_1)=
x_A'^\mu(s_1)-b^\nu\mathcal T_\nu'^\mu=
\xi_{A(1)}^\nu\der{\mathcal K_\nu^{(1)}}{p'_\mu}-b^\nu\mathcal T_\nu'^\mu \ ,
\end{gather}
from which it follows that
\begin{equation}\label{eq:-1}
-\delta\xi_{(1)}^\nu\der{\mathcal K_\nu^{(1)}}{p'_\mu}=\delta\xi_{(0)}^\nu\der{\mathcal K_\nu^{(0)}}{p'_\mu} \ .
\end{equation}
The fact that we are insisting only on a minimal requirement of translational invariance
is reflected also in the fact that our requirements are more general (weaker) than the
ones so far used for translational invariance in previous works on the relative-locality
framework. Our requirements (\ref{eq:}) and (\ref{eq:-1})
reproduce the ones enforced in Ref.~\cite{cortes}
upon  opting for boundary terms written in the
form $\displaystyle\bigoplus_{i=1}^{i=n}P^i_{in}-\bigoplus_{i=1}^{i=m}P^i_{out}$,
where $P^i_{in}$ are the ingoing momenta in a vertex and $P^i_{out}$ are the outgoing momenta. And our requirements (\ref{eq:}) and (\ref{eq:-1})
reproduce the strong translation transformations enforced in Ref.~\cite{anatomy},
by adopting $\delta\xi_{(1)}^\nu=\delta\xi_{(0)}^\nu=-b^\nu$, \emph{i.e.},  momentum independence of the $\xi^\mu$.

Let us next observe that from equation (\ref{eq:-1}) one has that
\begin{equation}
\delta\xi_{(0)}^\nu=
-\delta\xi_{(1)}^\sigma\der{\mathcal K_\sigma^{(1)}}{p'_\mu}\left(\der{\mathcal K_\nu^{(0)}}{p'_\mu}\right)^{-1} \ ,
\end{equation}
and using this in equation (\ref{eq:}) leads us to
\begin{equation}
\delta\xi_{(1)}^\sigma
\left[\der{\mathcal K_\sigma^{(1)}}{p_\rho}-\der{\mathcal K_\sigma^{(1)}}{p'_\mu}\left(\der{\mathcal K_\nu^{(0)}}{p'_\mu}\right)^{-1}\der{\mathcal K_\nu^{(0)}}{p_\rho}\right]=0  \ .
\end{equation}
Since $\delta\xi_{(1)}^\sigma\neq 0$ (in order for this to be a non-collapsed loop
the two observers must be distant) we conclude that
\begin{equation}\label{eq:-2}
\der{\mathcal K_\nu^{(1)}}{p_\rho}\left(\der{\mathcal K_\nu^{(1)}}{p'_\mu}\right)^{-1}-
\der{\mathcal K_\nu^{(0)}}{p_\rho}\left(\der{\mathcal K_\nu^{(0)}}{p'_\mu}\right)^{-1}=0 \ .
\end{equation}
This equation (\ref{eq:-2}) plays a pivotal role in our analysis since it shows that
an however weak requirement of translational invariance (required for relative locality
in a relativistic setup) imposes a restriction on the possible choices of boundary terms.
We shall now easily show that once the condition (\ref{eq:-2}) on boundary terms
  is enforced the causal loop is forbidden. We start by showing that for
  the boundary terms used in Ref.~\cite{linq} the condition (\ref{eq:-2})
   takes the shape of a condition on the momenta involved in the process, specifically, at leading order in  $\ell$,
   \begin{equation}\label{CLcovariance_condition}
\ell\delta_\mu^1\left[\delta_1^\rho\left(k'_0-q_0\right)+\delta_0^\rho\left(q'_1-k_1\right)\right]
=0 \ ,
\end{equation}
which implies that $k'_0=q_0+\mathcal O(\ell)$ and $q'_1=k_1+\mathcal O(\ell)$.
The fact that  the causal loop is forbidden can then be seen easily for example by
looking back at equation (\ref{eq:-5'}), now enforcing (\ref{CLcovariance_condition}):
one obtains
\begin{equation}
\Delta\tau u^\rho+\Delta\tau'\left[u'^\rho+u'^1\ell\left(\delta_0^\rho k_1-\delta_1^\rho k'_0\right)\right]=0 \ .
\label{jocextra}
\end{equation}
This excludes the causal loop for just the same reasons that, as observed earlier in this section, the causal loop is excluded in ordinary special relativity:
for $\rho=0$ equation (\ref{jocextra}),
\begin{equation}
\Delta\tau=-\Delta\tau'\frac{u'^0}{u^0}-\ell\Delta\tau'\left(\frac{u'^1k_1}{u^0}\right),
\end{equation}
 does not admit solutions with
positive $\Delta\tau$ and $\Delta\tau'$ and positive  zeroth component of  the two 4-velocities.
 This causal loop is indeed
  forbidden once a DSR-relativistic description, with associated relative locality, is enforced.

\section{M\"obius diagram and translational invariance}
\begin{figure}[h!]
\begin{center}
\includegraphics[scale=0.6]{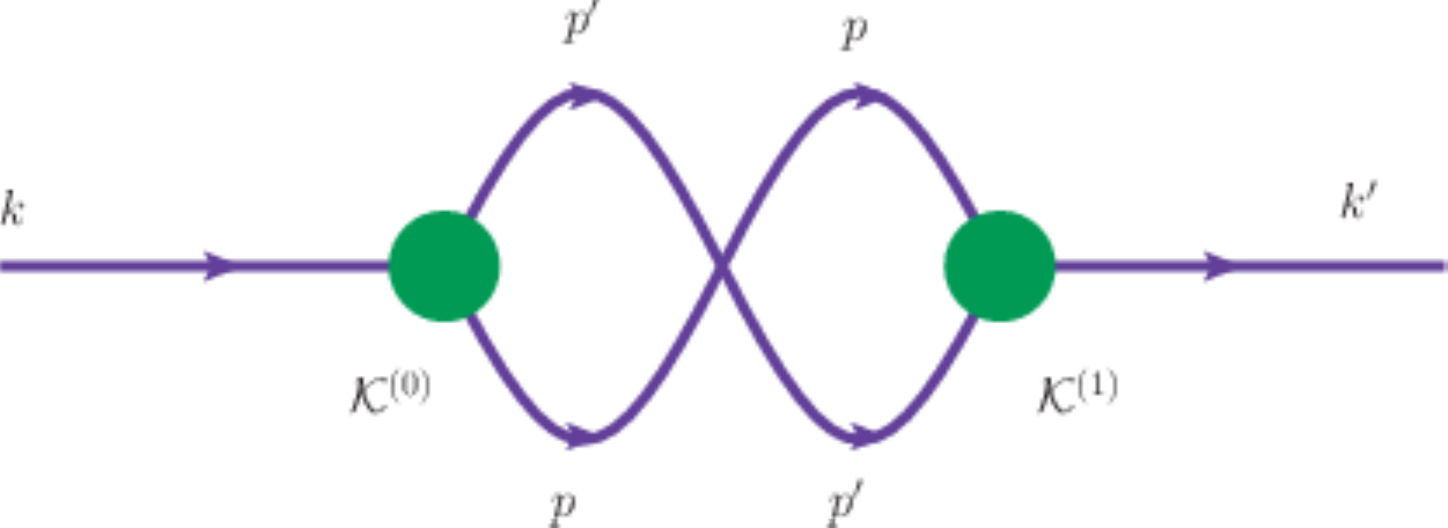}
\end{center}
\caption{\small We here show schematically two causally-connected events
that form a ``M\"obius diagram".
The laws of conservation at the two vertices
are setup in such a way that
the particle outgoing from
the first vertex has its momentum appearing
on the right-hand side of the composition law
and its momentum also appears on the left-hand side of the composition of momenta
at the second vertex.}
\label{loopandrb}
\end{figure}

Having shown that the
causal loop of Ref.\cite{linq}
is indeed allowed in generic
theories on curved momentum spaces, but is forbidden when a DSR-relativistic description, with associated relative spacetime locality, is enforced, we now proceed to the next announced task, which concerns the
diagram studied in Ref.\cite{andrb} as a possible source of violations of global momentum conservation.
Ref.\cite{andrb} considered theories on a curved momentum space, without
enforcing relative spacetime locality, and found that in general the diagram shown in our
Fig.\ref{loopandrb} can produce violations of global momentum conservation.
These violations take the shape~\cite{andrb} of $k' \neq k$, {\it i.e.} the momentum
incoming into the diagram is not equal to the momentum outgoing from the diagram.
Similarly to what we showed in the previous section for a causal loop, we shall find
that these violations of global momentum conservation from the diagram
in Fig.\ref{loopandrb} do not occur if one enforces a DSR-relativistic description, with associated relative spacetime locality.

The relative-locality-framework description of the diagram in Fig.\ref{loopandrb} is
obtained through the action
\begin{equation}
\begin{array}{lll}
\mathcal S
&=&
\displaystyle \int_{-\infty}^{s_0}ds\left(z^\mu\dot k_\mu+\mathcal N_k(\mathcal C_k-m_k^2)\right)
+\int_{s_1}^{+\infty}ds\left(z'^\mu\dot k'_\mu+\mathcal N_{k'}(\mathcal C_{k'}-m_{k'}^2)\right)+\\
&+&
\displaystyle
\int_{s_0}^{s_1}ds\left(x'^\mu\dot p'_\mu+\mathcal N_{p'}(\mathcal C_{p'}-m_{p'}^2)\right)+
\int_{s_0}^{s_1}ds\left(x^\mu\dot p_\mu+\mathcal N_p(\mathcal C_p-m_p^2)\right)+\\
&-&
\xi_{(0)}^\mu\mathcal K_\mu^{(0)}-\xi_{(1)}^\mu\mathcal K_\mu^{(1)},
\end{array}
\end{equation}
with
\begin{subequations}
\begin{align}
&\mathcal K_{\ \mu}^{(0)}=
\left(k \oplus\left(\ominus\left(p\oplus p'\right)\right)\right)_\mu \simeq
k_\mu-p_\mu-p'_\mu+\delta_\mu^1 \ell \left[p_1\left(k_0-p_0-p'_0\right)+p'_1\left(k_0-p'_0\right)\right],\label{MDLconservation1}\\
&\mathcal K_{\ \mu}^{(1)}=
\left(\left( p'\oplus p\right) \oplus \left( \ominus k'\right)\right)_\mu \simeq
p'_\mu+p_\mu-k'_\mu+\delta_\mu^1 \ell \left[k'_1 \left(p'_0+p_0-k_0'\right)-p'_0p_1\right].\label{MDLconservation2}
\end{align}
\end{subequations}
From the structure of (\ref{MDLconservation1})-(\ref{MDLconservation2}) it is clear why
we choose to label the diagram in Fig.\ref{loopandrb} as a ``M\"obius diagram":
the laws of conservation at the two vertices  use the
noncommutativity of the composition law in such a way
that the particle outgoing from
the first vertex with momentum appearing on the right-hand side of the composition law
enters the second vertex with momentum appearing on the left-hand side of the composition law. [Of course, the opposite applies to the other particle exchanged between the vertices]. If one then draws the diagram with the convention that the orientation
of pairs of legs entering/exiting a vertex
 consistently reflects the order in which
the momenta are composed, then the only way to draw the diagram makes it resemble a
M\"obius strip.

Evidently there is no room for such a structure when the momentum space has composition
law which is commutative. In particular there is no way to contemplate such a
M\"obius diagram in special relativity. But on our $\kappa$-momentum space
this structure is possible and its implications surely need to be studied.

Consistently with what we reported in the previous sections, our interest is into
understanding how the properties of the M\"obius diagram are affected
if one
enforces relative spacetime locality in DSR-relativistic theories
 on the
$\kappa$-momentum space. In particular, we want to show that $k' = k$ (no violation
of global momentum conservation).

As also already stressed above, relative spacetime locality in a relativistic
theory on curved momentum space necessarily requires at least a weak form of
translational invariance.
This insistence on at least the weakest possible notion of translational invariance
led us to find equations (\ref{eq:}) and (\ref{eq:-1}) for the causal loop,
and, as the interested reader can easily verify,
for the case of the M\"obius diagram it leads us to the equations
\begin{subequations}
\begin{gather}
\delta \xi^\nu_{(0)}\der{\mathcal K^{(0)}_\nu}{p_\mu}=
-\delta \xi^\nu_{(1)}\der{\mathcal K^{(1)}_\nu}{p_\mu},\\
\delta \xi^\nu_{(0)}\der{\mathcal K^{(0)}_\nu}{p'_\mu}=
-\delta \xi^\nu_{(1)}\der{\mathcal K^{(1)}_\nu}{p'_\mu}.
\end{gather}
\end{subequations}
These allow us to deduce that
\begin{equation}\label{transl_mobius}
\left[\der{\mathcal K^{(1)}_\sigma}{p_\mu}-\der{\mathcal K^{(1)}_\sigma}{p'_\rho}\left(\der{\mathcal K^{(0)}_\nu}{p'_\rho}\right)^{-1}\der{\mathcal K^{(0)}_\nu}{p_\mu}\right]=0.
\end{equation}
The implications of this equation
are best appreciated by exposing explicitly the momentum dependence
of the terms appearing in (\ref{transl_mobius}):
\begin{subequations}\label{derKandrb}
\begin{gather}
\der{\mathcal K^{(1)}_\sigma}{p_\mu}=
\delta_\sigma^\mu +\ell \delta_\sigma^1  \left(\delta_0^\mu k'_1-\delta_1^\mu p'_0\right),\\
\der{\mathcal K^{(1)}_\sigma}{ p'_\rho}=
\delta_\sigma^\rho+\ell \delta_\sigma^1\delta_0^\rho\left(k'_1-p_1\right),\\
\left(\der{\mathcal K^{(0)}_\nu}{p'_\rho}\right)^{-1}=
-\delta_\rho^\nu-\ell \delta_\rho^1\left[\delta^\nu_1\left(k_0-p'_0\right)-\delta_0^\nu\left( p_1+p'_1\right)\right],\\
\der{\mathcal K^{(0)}_\nu}{p_\mu}=
-\delta^\mu_\nu-\ell \delta^\mu_0\delta^1_\nu p_1.
\end{gather}
\end{subequations}
These allow us to conclude that
from (\ref{transl_mobius}) it follows that
\begin{equation}\label{MDLcovariance_condition}
\ell \left[\delta_1^\mu k_0-\delta_0^\mu \left(p_1+p'_1\right)\right]=0.
\end{equation}
Using this result in combination with the conservation laws $\mathcal K^{(0)}_{\mu}=0$
and $\mathcal K^{(1)}_{\mu}=0$ one can easily establish that
\begin{equation}\label{MDLp+p'}
p_\mu+p'_\mu=0+\mathcal O(\ell)~,
\end{equation}
and one can also rewrite those conservation laws as follows
\begin{gather}
0=k_\mu-p_\mu-p'_\mu-\delta_\mu^1\ell p'_1p'_0,\label{consJOCa}\\
0=p'_\mu+p_\mu-k'_\mu- \delta_\mu^1\ell p'_0p_1 \, \, .
\label{consJOCb}
\end{gather}
Summing these (\ref{consJOCa}) and (\ref{consJOCb}), also using
  (\ref{MDLp+p'}), we get to the sought result
\begin{equation}
k_\mu=k'_\mu+\mathcal O(\ell^2) \,\, ,
\end{equation}
showing that indeed by insisting on a having a DSR-relativistic picture, with associated relative spacetime locality, one finds  no global violation of momentum conservation (at least
at leading order in $\ell$, which is the level of accuracy we are here pursuing).

\section{Combinations of M\"obius diagrams and implications for building a quantum theory}

In the previous section we reported results suggesting that when theories  are (DSR-)relativistic, with the associated relativity of spacetime locality,
momentum is globally conserved and there is no violation of causality.
It should be noticed  that the objective of  enforcing relative spacetime locality led us to introduce some restrictions on the choice of boundary terms,  particularly for causally-connected interactions. The relevant class of theories has been studied so far only within the confines of classical mechanics, and therefore such prescriptions concerning boundary terms are meaningful and unproblematic (they can indeed be enforced by principle, as a postulate). The quantum version of the theories we here considered is still not known, but if one tries to imagine which shape it might take it seems that enforcing the principle of relative locality  in a quantum theory might be very challenging: think in particular  of quantum field theories formulated in terms of a generating functional, where all such prescriptions are usually introduced by a single specification of the generating functional. While we do not have anything to report on this point which would address directly the challenges for the construction of such quantum theories, we find it worthy to provide evidence of the fact that combinations of diagrams on curved momentum space might have less anomalous properties, even without enforcing relative locality, than single diagrams.

In an appropriate sense we are attempting to provide first elements
in support of a picture which we conjecture ultimately
might be somewhat analogous to what happens, for example,
in the analysis of the gauge invariance of the first contribution to the matrix element of the
Compton scattering $e^-+\gamma\rightarrow e^-+\gamma$ in standard QED. In fact, in that case there are only two Feynman diagrams that need to be taken into account and the matrix element is given by
\begin{equation}
\mathcal{M}_{fi}=(-ie)^2\left(\bar{u}_{p'}\slashed{\epsilon}(q')\frac{i}{\slashed{p}+\slashed{q}-m}\slashed{\epsilon} (q)u_p+\bar{u}_{p'}\slashed{\epsilon}(q)\frac{i}{\slashed{p}-\slashed{q}'-m}\slashed{\epsilon}(q')u_p \right),
\label{jocmm}
\end{equation}
where $p$ and $q$ are the momenta of the electron and the photon respectively in the initial state,  $p'$ and $q'$ are the momenta of the electron and the photon respectively in the final state, $u_p$ and $\bar{u}_p$ are Dirac spinors and $\epsilon_\mu$ is the photon polarization 4-vector. For a free photon described in the Lorentz gauge by a  plane wave $A_\mu(x) \propto \epsilon_\mu(k)e^{\pm ik_\nu x^\nu}$ the gauge transformation $A_\mu^\Lambda(x)=A_\mu(x) +\partial_\mu \Lambda(x)$ with $\Lambda(x)=\tilde{\Lambda}(k)e^{\pm ik_\nu x^\nu}$ corresponds to a transformation of the polarization 4-vector  $\epsilon_\mu ^\Lambda(k)=\epsilon_\mu(k)\pm ik_\mu \tilde{\Lambda}(k)$.
Equipped with these observations one can easily see that
 the two terms in (\ref{jocmm}) are not individually gauge
 invariant, but their combination is gauge invariant.

We are not going to provide conclusive evidence that a similar mechanism is at work for causality and global momentum conservation in theories on curved momentum space (it would be impossible without knowing how to formulate such a quantum theory), but it may be nonetheless interesting that we can find some points of intuitive connection with stories such as that  of gauge invariance for Compton scattering.

For definiteness and simplicity we focus on the case of M\"obius diagrams.
In the previous section we analyzed a M\"obius diagram using the choice of boundary terms adopted in Ref.~\cite{andrb}  since the appreciation of the presence of a challenge due to M\"obius diagrams originated from the study reported in Ref.~\cite{andrb}. In this section we look beyond the realm of considerations offered in  Ref.~\cite{andrb}, so we go back to our
preferred criterion for the choice of boundary conditions, the one
first advocated in Ref.\cite{anatomy}, which allows us to streamline the derivations.
 So we consider the M\"obius diagram
 by adopting the following prescription for the boundary terms:
\begin{equation}\label{trevMob}
\begin{split}
\mathcal K^{(0)}_\mu=k_\mu-(p\oplus p')_\mu\simeq k_\mu-p_\mu-p'_\mu+\ell\delta_\mu^1p_0p'_1,\\
\mathcal K^{(1)}_\mu=(p'\oplus p)_\mu-k'_\mu\simeq p'_\mu+p_\mu-k_\mu-\ell\delta_\mu^1p'_0p_1.
\end{split}
\end{equation}
From the conservation of four-momentum  at each vertex $\mathcal K^{(0)}_\mu=0$, $\mathcal K^{(1)}_\mu=0$ we get
\begin{equation}\label{deltaMomentaMobius1}
k_\mu-k'_\mu=\ell\delta_\mu^1(p'_0p_1-p_0p'_1)=\ell\delta_\mu^1(\frac{m_{p}^2p'_1}{2p_1}-\frac{m_{p'}^2p_1}{2p'_1})\equiv \ell \delta_\mu^1\Delta ,
\end{equation}
where, since we are considering particles of energy-momentum $\ell^{-1}\gg p_\mu\gg m$, from the on-shell relation (\ref{geomassJ}) we expressed the energy of the particles as $p_0=\sqrt{p_1^2+m^2}+\frac{\ell p_1^2}{2}\approx |p_1|+\frac{m^2}{2|p_1|}+\frac{\ell p_1^2}{2}$.

At this point we must stress that evidently this is not the only way to have a M\"obius diagram, since we can interchange the prescription for which  particle  enters the composition law for the first event on the right side of the composition law (then entering the second event on the left side of the composition law). This alternative possibility (which is the only
other possibility allowed within the prescriptions of Ref.~\cite{anatomy})
 is characterized by boundary terms of the form
\begin{equation}\label{trevMob2}
\begin{split}
\tilde{\mathcal K}^{(0)}_\mu=k_\mu-(p'\oplus p)_\mu\simeq k_\mu-p'_\mu-p_\mu+\ell\delta_\mu^1p'_0p_1,\\
\tilde{\mathcal K}^{(1)}_\mu=(p\oplus p')_\mu-k'_\mu\simeq p'_\mu+p_\mu-k'_\mu-\ell\delta_\mu^1p_0p'_1.
\end{split}
\end{equation}
Then the condition one obtains in place of (\ref{deltaMomentaMobius1}) is
\begin{equation}\label{deltaMomentaMobius2}
k_\mu-k'_\mu=-\ell\delta_\mu^1\Delta.
\end{equation}

Of course, in light of what we established in the previous section, both
of these M\"obius diagrams must be excluded if one enforces the principle of relative spacetime locality. But it is interesting to notice that if we were to allow these M\"obius diagrams the violation of global momentum conservation
produced by one of them,  (\ref{deltaMomentaMobius1}), is exactly opposite to the one produced by the other one, (\ref{deltaMomentaMobius2}). In a quantum-field
theory version of the classical theories we here analyzed  one might have to include together these opposite contributions, in which case we conjecture that the net result would not be some systematic
prediction of violation of global momentum conservation, but rather something
of the sort rendering global momentum still conserved but fuzzy.

Going back to the classical-mechanics version of these theories it is amusing to notice
that  a chain composed of two M\"obius diagrams, one of type (\ref{deltaMomentaMobius1})
and one of type (\ref{deltaMomentaMobius2}), would have as net result no violation of global momentum conservation.

\section{Summary and outlook}
The study of Planck-scale-curved momentum spaces is presently at a point of balance between growing supporting evidence and concerns about its consistency with established experimental facts. On one side, as stressed in our opening remarks, the list of quantum-gravity approaches where these momentum-space-curvature effects are encountered keeps growing, and interest in this possibility is also rooted in some opportunities for a dedicated phenomenological programme with Planck-scale sensitivity\cite{principle,gacLRR}.
On the other hand it is increasingly clear that in general theories on curved momentum space may violate several apparently robust aspects of our current description of the laws of physics, including relativistic invariance, locality, causality and global momentum conservation. We here contributed to the characterization of how severe these challenges can be for generic theories on curved momentum spaces, but we also reported results suggesting that when the theory is formulated (DSR-)relativistically, with the associated relativity of spacetime locality,
momentum is globally conserved and there is no violation of causality. It seems then that (at least in this first stages of exploration) it might be appropriate to restrict the focus of research on curved momentum space on this subclass with more conventional properties, which one should expect when the momentum space is maximally symmetric.

It should be noticed that here (just like in Refs.~\cite{linq,andrb})
we only considered the simplest chain of events that could have led to violations of causality and global momentum conservation. That already involved some significant technical challenges, but does not suffice to show that in general causality and glabal momentum conservation are ensured when these theories are  formulated (DSR-)relativistically, with the associated relativity of spacetime locality. The fact that the violations are in general  present for the simple chains of events we analyzed but disappear
when relative locality is enforced
is surely of strong encouragement but does not represent a general result.

Of course, the main challenge on the way toward greater maturity for this novel research programme is the development of a quantum-field-theory version. As we were in the final stages of the writeup of this manuscript a general framework for introducing such quantum field theories was proposed
in Ref.~\cite{freidelFT}. While presently this proposal
appears to be still at too early and too formal a stage
of development for addressing the challenges that were here of interest, it is legitimate to hope that, as its understanding deepens, a consistent quantum picture of causality and momentum conservation with curved momentum spaces will arise.


\begin{thebibliography}{50}

\bibitem{majidCURVATURE}
S.~Majid, arXiv:hep-th/0006166, Lect. Notes Phys. 541 (2000) 227.

\bibitem{dsr1}
G.~Amelino-Camelia, arXiv:gr-qc/0012051, Int. J. Mod. Phys. D11 (2002) 35

\bibitem{dsr2}
G.~Amelino-Camelia, arXiv:hep-th/0012238, Phys. Lett. B510 (2001) 255.

\bibitem{jurekDSMOMENTUM}
J.~Kowalski-Glikman, arXiv:hep-th/0207279, Phys. Lett. B547 (2002) 291.

\bibitem{girelliCURVATURE}
F.~Girelli, E.R.~Livine, arXiv:gr-qc/0412079, Braz. J. Phys. 35 (2005) 432.

\bibitem{schullerCURVATURE}
D.~Ratzel, S.~Rivera and F.P.~Schuller, Phys. Rev. D83 (2011) 044047.

\bibitem{changMINIC}
L.N.~Chang, Z.~Lewis, D.~Minic and T.~Takeuchi, arXiv:1106.0068, Adv. High Energy Phys. 2011 (2011) 493514.


\bibitem{principle}
G.~Amelino-Camelia, L.~Freidel, J.~Kowalski-Glikman, L.~Smolin,
arXiv:1101.0931,
Phys. Rev. D84 (2011) 084010.

\bibitem{gacmaj}
G.~Amelino-Camelia, S.~Majid, arXiv:hep-th/9907110, Int. J. Mod. Phys. A15 (2000) 4301.


\bibitem{rovelliLRR}
C.~Rovelli, Living Rev. Relativity 11, (2008) 5.

\bibitem{leeCURVEDMOMENTUM}
L.~Smolin, Lect. Notes Phys. 669 (2005) 363.


\bibitem{dsr3FREIDLIVINE} L.~Freidel, E. R. Livine, arXiv:hep-th/0512113, Phys. Rev. Lett. 96 (2006) 221301.

\bibitem{matschull} H.~-J.~Matschull, M.~Welling, arXiv:gr-qc/9708054, Class. Quant. Grav. 15 (1998) 2981.

\bibitem{BaisMullerSchroers} F.~A.~Bais, N.~M.~Muller, B.~J.~Schroers, arXiv:hep-th/0205021, Nucl. Phys. B640 (2002) 3.

\bibitem{bernschr2012}
    P.K. Osei, B.J. Schroers
    J. Math. Phys. 53 (2012) 073510.

\bibitem{stefanoCQG2013} G.~Amelino-Camelia, M.~Arzano, S.~Bianco, R.~J.~Buonocore,  arXiv:1210.7834, Class.Quant.Grav. 30 (2013) 065012.

\bibitem{GiuliaFlavio}
G.~Gubitosi, F.~Mercati, arXiv:1106.5710, Class. Quant. Grav. 30 (2013) 145002.


\bibitem{GACarXiv11105081}
G.~Amelino-Camelia, arXiv:1110.5081, Phys. Rev. D85 (2012) 084034.


\bibitem{cortes}
J.~M.~Carmona, J.~L.~Cortes, D.~Mazon, F.~Mercati, arXiv:1107.0939,
Phys. Rev. D84 (2011) 085010.

\bibitem{jurekDSRfirst}
J.~Kowalski-Glikman, arXiv:hep-th/0102098, Phys. Lett. A286 (2001) 391-394.

\bibitem{leejoaoPRDdsr}
L.~Smolin and J. Magueijo, arXiv:gr-qc/0207085, Phys. Rev. D67 (2003) 044017.

\bibitem{leejoaoCQGrainbow}
L.~Smolin and J. Magueijo, arXiv:gr-qc/0305055, Class. Quant. Grav. 21 (2004) 1725-1736.

\bibitem{jurekDSRreview}
J.~Kowalski-Glikman, arXiv:hep-th/0405273, Lect. Notes Phys. 669 (2005) 131-159.

\bibitem{gacSYMMETRYreview}
G.~Amelino-Camelia, arXiv:1003.3942, Symmetry 2 (2010) 230-271.

\bibitem{whataboutbob}
G.~Amelino-Camelia, M.~Matassa, F.~Mercati, G.~Rosati, arXiv:1006.2126, Phys. Rev. Lett. 106 (2011) 071301.


\bibitem{linq}
L.~-Q.~Chen, arXiv:1212.5233,  Phys. Rev. D88 (2013) 024052.

\bibitem{andrb}
A.~Banburski, arXiv:1305.7289, Phys. Rev. D88 (2013) 076012.

\bibitem{anatomy}
G.~Amelino-Camelia, M.~Arzano, J.~Kowalski-Glikman, G.~Rosati and G.~Trevisan, arXiv:1107.1724,
Class. Quant. Grav.  {29} (2012) 075007.

\bibitem{majrue}
S.~Majid, H.~Ruegg, arXiv:hep-th/9405107, Phys. Lett. B334 (1994) 348-354.

\bibitem{lukieANNALS}
J.~Lukierski, H.~Ruegg, W.J.~Zakrzewski, arXiv:hep-th/9312153, Annals Phys. 243 (1995) 90.

\bibitem{gacLRR}
 G.~Amelino-Camelia, arXiv:0806.0339,
  Living Rev.\ Rel.\  {16} (2013) 5.

\bibitem{freidelFT}
L. Freidel and T. Rempel,
arXiv:1312.3674.

\end{thebibliography}
\end{document}